\documentclass{IEEEtran}
\usepackage[utf8]{inputenc}
\usepackage{graphicx} % Required for inserting images
\usepackage[american,siunitx]{circuitikz}
\usepackage{pgfplots}
\usepgfplotslibrary{external,colormaps}
\pgfplotsset{compat=1.10}

\usepackage{amsmath,amssymb,amsfonts}
\usepackage{graphicx}
\usepackage{cite}
\usepackage{hyperref}

\usepackage{caption}
\usepackage{subcaption}

\usepackage{balance}

\usepackage{multirow}

\usepackage{fancyvrb}
\usepackage{enumitem}

\usepackage{float}
\usepackage{microtype}  
\usepackage{listings}
\usepackage{xcolor}
\usepackage[scaled=.92]{inconsolata}  % optional: nicer mono font
\usepackage{mdframed}

\definecolor{PromptGrey}{RGB}{110,115,120}
\definecolor{ReplyTeal}{RGB}{28,105,110}
\definecolor{ReasoningOrange}{RGB}{255, 150, 79}

\mdfdefinestyle{chatbase}{%
  leftline=true, topline=false, rightline=false, bottomline=false,
  linewidth=0.8pt,
  innerleftmargin=6pt, innerrightmargin=6pt,
  innertopmargin=4pt,  innerbottommargin=4pt,
  roundcorner=1pt,
  font=\footnotesize\ttfamily,
  skipabove=\baselineskip, skipbelow=\baselineskip,
  frametitlerule=false,
  frametitleaboveskip=-\dimexpr\baselineskip-4pt\relax,
  frametitlebelowskip=4pt,
}

\mdfdefinestyle{promptstyle}{%
  style=chatbase,
  linecolor=PromptGrey,
  backgroundcolor=PromptGrey!6,
  frametitlefont=\ttfamily\bfseries\color{PromptGrey},
  frametitlebackgroundcolor=PromptGrey!6,
}

\newmdenv[
  style=promptstyle,
  frametitle={Prompt},
]{prompt}

\mdfdefinestyle{responsestyle}{%
  style=chatbase,
  linecolor=ReplyTeal,
  backgroundcolor=ReplyTeal!5,
  frametitlefont=\ttfamily\bfseries\color{ReplyTeal},
  frametitlebackgroundcolor=ReplyTeal!5,
}

\newmdenv[
  style=responsestyle,
  frametitle={Response},
]{response}

\mdfdefinestyle{reasoningstyle}{%
  style=chatbase,
  linecolor=ReasoningOrange,
  backgroundcolor=ReasoningOrange!5,
  frametitlefont=\ttfamily\bfseries\color{ReasoningOrange},
  frametitlebackgroundcolor=ReasoningOrange!5,
}

\newmdenv[
  style=reasoningstyle,
  frametitle={Reasoning},
]{reasoning}

\definecolor{codegreen}{rgb}{0,0.6,0}
\definecolor{codegray}{rgb}{0.5,0.5,0.5}
\definecolor{codepurple}{rgb}{0.58,0,0.82}
\definecolor{backcolour}{rgb}{0.95,0.95,0.92}

\lstdefinestyle{mystyle}{
    backgroundcolor=\color{ReasoningOrange!5},   
    commentstyle=\color{codegreen},
    keywordstyle=\color{magenta},
    numberstyle=\tiny\color{codegray},
    stringstyle=\color{codepurple},
    basicstyle=\ttfamily\footnotesize,
    numbers=left,                    % show line numbers on the left
    numbersep=8pt,                   % gap between numbers and code
    xleftmargin=1.5em,               % indent everything (numbers+code) from left column edge
    framexleftmargin=1em,            % extra space if you also draw a frame
    breaklines=true,
    breakatwhitespace=false,
    breakindent=1em,                 % indent for wrapped lines
    keepspaces=true,
    showstringspaces=false,
    tabsize=2,
    captionpos=b
}

\usepackage{tikz}
\usetikzlibrary{arrows,shapes,positioning}
\usepackage{ifthen}
\usetikzlibrary{calc,intersections,through,backgrounds}
\usetikzlibrary{decorations.pathmorphing}
\usetikzlibrary{matrix}

\tikzset{sin v source/.style={
  circle,
  draw,
  append after command={
    \pgfextra{
    \draw
      ($(\tikzlastnode.center)!0.5!(\tikzlastnode.west)$)
       arc[start angle=180,end angle=0,radius=0.425ex] 
      (\tikzlastnode.center)
       arc[start angle=180,end angle=360,radius=0.425ex]
      ($(\tikzlastnode.center)!0.5!(\tikzlastnode.east)$) 
    ;
    }
  },
  scale=1.5,
 }
}

\tikzset{
  declare function={% in case of CVS which switches the arguments of atan2
    atan3(\a,\b)=ifthenelse(atan2(0,1)==90, atan2(\a,\b), atan2(\b,\a));},
  kinky cross radius/.initial=+.125cm,
  @kinky cross/.initial=+, kinky crosses/.is choice,
  kinky crosses/left/.style={@kinky cross=-},kinky crosses/right/.style={@kinky cross=+},
  kinky cross/.style args={(#1)--(#2)}{
    to path={
      let \p{@kc@}=($(\tikztotarget)-(\tikztostart)$),
          \n{@kc@}={atan3(\p{@kc@})+180} in
      -- ($(intersection of \tikztostart--{\tikztotarget} and #1--#2)!%
             \pgfkeysvalueof{/tikz/kinky cross radius}!(\tikztostart)$)
      arc [ radius     =\pgfkeysvalueof{/tikz/kinky cross radius},
            start angle=\n{@kc@},
            delta angle=\pgfkeysvalueof{/tikz/@kinky cross}180 ]
      -- (\tikztotarget)}}}

\newcommand{\wyeload}[8]{%
  \coordinate (#1) at #2;

  \ifthenelse{\equal{#5}{}}%
  { % No voltage labels, so don't draw polarity symbols
      \draw (#1) to[R,l=\mbox{\ensuremath{#3}},*-*] +({0+90}:3) coordinate (#1_a) node[anchor=south]{}; %$V_a$
  }
  { % Draw voltage labels and polarity symbols
    \draw (#1) to[R,l=\mbox{\ensuremath{#3}},*-*,v_<=\mbox{\ensuremath{#5}}] +({0+90}:3) coordinate (#1_a) node[anchor=south]{}; %$V_a$
  }
  
  \ifthenelse{\equal{#6}{}}%
  { % No voltage labels, so don't draw polarity symbols
      \draw (#1) to[R,l_=\mbox{\ensuremath{#3}},*-*] +({-120+90}:3) coordinate (#1_b) node[anchor={120}]{}; %$V_b$
  }
  { % Draw voltage labels and polarity symbols
    \draw (#1) to[R,l_=\mbox{\ensuremath{#3}},*-*,v^<=\mbox{\ensuremath{#6}}] +({-120+90}:3) coordinate (#1_b) node[anchor={120}]{}; %$V_b$
  }

  \ifthenelse{\equal{#7}{}}%
  { % No voltage labels, so don't draw polarity symbols
      \draw (#1) to[R,l=\mbox{\ensuremath{#3}},*-*] +({120+90}:3) coordinate (#1_c) node[anchor={-120}]{}; %$V_c$
  }
  { % Draw voltage labels and polarity symbols
    \draw (#1) to[R,l=\mbox{\ensuremath{#3}},*-*,v_<=\mbox{\ensuremath{#7}}] +({120+90}:3) coordinate (#1_c) node[anchor={-120}]{}; %$V_c$
  }
  
  \draw (#1) +(0.22,-0.1) node[anchor=south]{\ensuremath{#8}};

  \ifthenelse{\equal{#4}{}}%
  {% No impedance to ground, so do nothing
  }
  {% Draw impedance to ground
    \draw (#1) to[R,l=\mbox{\ensuremath{#4}},*-,/tikz/circuitikz/bipoles/length=1cm] +(0,-1.5);
    \draw ($(#1)+(0,-1.15)$) node[ground]{};
  }
}

\newcommand{\deltaload}[6]{%
  \coordinate (#1) at #2;
  \coordinate (#1_a) at ($(#1)+({0+90}:3)$);
  \coordinate (#1_b) at ($(#1)+({-120+90}:3)$);
  \coordinate (#1_c) at ($(#1)+({120+90}:3)$);

  \ifthenelse{\equal{#4}{}}%
  { % No voltage labels, so don't draw polarity symbols
    \draw (#1_b) to[R,l_=\mbox{\ensuremath{#3}},*-*] (#1_a);
  }
  { % Draw voltage labels and polarity symbols
    \draw (#1_b) to[R,l_=\mbox{\ensuremath{#3}},*-*,v^<=\mbox{\ensuremath{#4}}] (#1_a);
  }

  \ifthenelse{\equal{#5}{}}%
  { % No voltage labels, so don't draw polarity symbols
    \draw (#1_c) to[R,l_=\mbox{\ensuremath{#3}},*-*] (#1_b);
  }
  { % Draw voltage labels and polarity symbols
    \draw (#1_c) to[R,l_=\mbox{\ensuremath{#3}},*-*,v^<=\mbox{\ensuremath{#5}}] (#1_b);
  }

  \ifthenelse{\equal{#6}{}}%
  { % No voltage labels, so don't draw polarity symbols
    \draw (#1_a) to[R,l_=\mbox{\ensuremath{#3}},*-*] (#1_c);
  }
  { % Draw voltage labels and polarity symbols
    \draw (#1_a) to[R,l_=\mbox{\ensuremath{#3}},*-*,v^<=\mbox{\ensuremath{#6}}] (#1_c);
  }
}

\newcommand{\labelcurrentsWye}[5]{%
    \ifthenelse{\equal{#5}{out}}%
      {% Draw currents flowing out of the neutral
        \draw (#1) to[open,i^=\mbox{\ensuremath{#2}}] (#1_a);
        \draw (#1) to[open,i_=\mbox{\ensuremath{#3}}] (#1_b);
        \draw (#1) to[open,i^=\mbox{\ensuremath{#4}}] (#1_c);
      }
      {% Draw currents flowing into the neutral
        \draw (#1_a) to[open,i>_=\mbox{\ensuremath{#2}}] (#1);
        \draw (#1_b) to[open,i>^=\mbox{\ensuremath{#3}}] (#1);
        \draw (#1_c) to[open,i>_=\mbox{\ensuremath{#4}}] (#1);
      }
}

\newcommand{\labelcurrentsDelta}[5]{%
    \ifthenelse{\equal{#5}{AB}}%
      {% Draw currents flowing from phase A->B, B->C, and C->A
        \draw (#1_a) to[open,i>^=\mbox{\ensuremath{#2}}] (#1_b);
        \draw (#1_b) to[open,i>^=\mbox{\ensuremath{#3}}] (#1_c);
        \draw (#1_c) to[open,i>^=\mbox{\ensuremath{#4}}] (#1_a);
      }
      {% Draw currents flowing into the neutral
        \draw (#1_a) to[open,i<_=\mbox{\ensuremath{#2}}] (#1_b);
        \draw (#1_b) to[open,i<_=\mbox{\ensuremath{#3}}] (#1_c);
        \draw (#1_c) to[open,i<_=\mbox{\ensuremath{#4}}] (#1_a);
      }
}

\newcommand{\labelLLvoltages}[5]{%
    \ifthenelse{\equal{#5}{AB}}%
      {% Draw voltages with polarity from phase AB, BC, and CA
        \draw ($(#1_a)+(0.5,0.25)$) to[open,v^=\mbox{\ensuremath{#2}}] ($(#1_b)+(0.5,0.25)$);
        \draw ($(#1_b)+(0.3,-0.5)$) to[open,v^=\mbox{\ensuremath{#3}}] ($(#1_c)+(-0.3,-0.5)$);
        \draw ($(#1_c)+(-0.3,0.5)$) to[open,v^=\mbox{\ensuremath{#4}}] ($(#1_a)+(-0.5,0.25)$);
      }
      {% Draw voltages with polarity from phase BA, CB, and AC
        \draw ($(#1_b)+(0.5,0.25)$) to[open,v_=\mbox{\ensuremath{#2}}] ($(#1_a)+(0.5,0.25)$);
        \draw ($(#1_c)+(-0.3,-0.5)$) to[open,v_=\mbox{\ensuremath{#3}}] ($(#1_b)+(0.3,-0.5)$);
        \draw ($(#1_a)+(-0.5,0.25)$) to[open,v_=\mbox{\ensuremath{#4}}] ($(#1_c)+(-0.3,0.5)$);
      }
}

\newcommand{\connectcomponents}[9]{%

    \coordinate (#1#2_int1) at ($(#1_b)-(0,#3)$);
    \coordinate (#1#2_int2) at ($(#2_b)-(0,#3)$);
    \coordinate (#1#2_int3) at ($(#2_c)-(0,#3*1.5)$);

    \path [name path=#1#2path1] (#1#2_int1) -- (#1#2_int2);
    \path [name path=#1#2path2] (#1#2_int3) -- (#2_c);
    \path [name intersections={of=#1#2path1 and #1#2path2,by=#1#2cross}];

    \ifthenelse{\equal{#4}{}}%
    {% Don't draw series impedance
        \ifthenelse{\equal{#5}{}}%
        {% Don't draw the line current arrow
            \draw (#1_a) -- ($(#1_a)+(0,#9)$) to[short,-] ($(#2_a)+(0,#9)$) -- (#2_a); 
            \draw (#1_b) -- (#1#2_int1) to[short,-] (#1#2_int2) -- (#2_b);
            \draw (#1_b) -- (#1#2_int1) to[short,-] (#1#2_int2) -- (#2_b);
            \draw (#1_c) -- ($(#1_c)-(0,#3*1.5)$) to[short,-] (#1#2_int3) to[kinky cross=(#1#2cross)--(#1#2_int1), kinky crosses=left] (#2_c);        
        }
        {% Draw the line current arrow and label
            \draw (#1_a) -- ($(#1_a)+(0,#9)$) to[short,-,i_=\mbox{\ensuremath{#5}}] ($(#2_a)+(0,#9)$) -- (#2_a); 
            \draw (#1_b) -- (#1#2_int1) to[short,-,i_=\mbox{\ensuremath{#6}}] (#1#2_int2) -- (#2_b);
            \draw (#1_b) -- (#1#2_int1) to[short,-] (#1#2_int2) -- (#2_b);
            \draw (#1_c) -- ($(#1_c)-(0,#3*1.5)$) to[short,-,i_=\mbox{\ensuremath{#7}}] (#1#2_int3) to[kinky cross=(#1#2cross)--(#1#2_int1), kinky crosses=left] (#2_c);
        }
    }
    {% Draw series impedance
        \ifthenelse{\equal{#5}{}}%
        {% Don't draw the line current arrow
            \draw (#1_a) -- ($(#1_a)+(0,#9)$) to[R,l_=\mbox{\ensuremath{#4}},-] ($(#2_a)+(0,#9)$) -- (#2_a); 
            \draw (#1_b) -- (#1#2_int1) to[R,l_=\mbox{\ensuremath{#4}},-] (#1#2_int2) -- (#2_b);
            \draw (#1_b) -- (#1#2_int1) to[R,l_=\mbox{\ensuremath{#4}},-] (#1#2_int2) -- (#2_b);
            \draw (#1_c) -- ($(#1_c)-(0,#3*1.5)$) to[R,l_=\mbox{\ensuremath{#4}},-] (#1#2_int3) to[kinky cross=(#1#2cross)--(#1#2_int1), kinky crosses=left] (#2_c);
        }
        {% Draw the line current arrow and label
            \draw (#1_a) -- ($(#1_a)+(0,#9)$) to[R,l_=\mbox{\ensuremath{#4}},-,i=\mbox{\ensuremath{#5}}] ($(#2_a)+(0,#9)$) -- (#2_a); 
            \draw (#1_b) -- (#1#2_int1) to[R,l_=\mbox{\ensuremath{#4}},-,i_=\mbox{\ensuremath{#6}}] (#1#2_int2) -- (#2_b);
            \draw (#1_b) -- (#1#2_int1) to[R,l_=\mbox{\ensuremath{#4}},-] (#1#2_int2) -- (#2_b);
            \draw (#1_c) -- ($(#1_c)-(0,#3*1.5)$) to[R,l_=\mbox{\ensuremath{#4}},-,i_=\mbox{\ensuremath{#7}}] (#1#2_int3) to[kinky cross=(#1#2cross)--(#1#2_int1), kinky crosses=left] (#2_c);
        }
    }

    \ifthenelse{\equal{#8}{}}%
      {% No neutral impedance connection, so nothing to do here.
      }
      {% Draw impedance connecting neutrals
        \draw (#1) -- ++(0.5,0) -- ++(0,0.5) to[R,l_=\mbox{\ensuremath{#8}}] ($(#2)+(-0.3,0.5)$) -- ++(0,-0.5) -- ++(0.25,0);
      }
}

\newcommand{\transformer}[9]{%

    \coordinate (#1) at #2;
    \coordinate (#1_left) at ($(#1)+(-0.75,0)$);
    \coordinate (#1_right) at ($(#1)+(0.75,0)$);

    \draw (#1_left) -- +(0.6,0);
    \draw (#1_right) -- +(-0.6,0);

    \draw ($(#1)+(-0.2,0.3)$) arc (-90:90:0.15);
    \draw ($(#1)+(-0.2,0)$) arc (-90:90:0.15);
    \draw ($(#1)+(-0.2,-0.3)$) arc (-90:90:0.15);
    \draw ($(#1)+(-0.2,-0.6)$) arc (-90:90:0.15);

    \draw ($(#1)+(0.2,0.6)$) arc (90:270:0.15);
    \draw ($(#1)+(0.2,0.3)$) arc (90:270:0.15);
    \draw ($(#1)+(0.2,0)$) arc (90:270:0.15);
    \draw ($(#1)+(0.2,-0.3)$) arc (90:270:0.15);
    
    \ifthenelse{\equal{#3}{delta}}%
      {% Label left side with a delta connection
        \draw ($(#1)+(-0.6,-0.45)$) node{\Large $\Delta$};
      }
      {% Label left side with a wye connection
        \draw ($(#1)+(-0.6,-0.4)$) to[short, -] ++(-90:0.3) node[left]{};
        \draw ($(#1)+(-0.6,-0.4)$) to[short, -] ++(30:0.3)  node[above]{};
        \draw ($(#1)+(-0.6,-0.4)$) to[short, -] ++(150:0.3) node[left]{};
        \ifthenelse{\equal{#3}{wye}}%
        {% Draw an ungrounded wye (nothing additional required)
        }%
        {\ifthenelse{\equal{#3}{0}}%
            {% Draw a wye with a neutral shorted to ground
                \draw [/tikz/circuitikz/bipoles/length=0.6cm] ($(#1)+(-0.6,-0.4)$) to[short,-] ++(-0.3,0) to[short, -] ++(0,-0.1) node[ground]{};
            }
            {% Draw a wye with a neutral grounded through the specified impedance
                \draw [/tikz/circuitikz/bipoles/length=0.6cm] ($(#1)+(-0.6,-0.4)$) to[short,-] ++(-0.3,0) to[short, -] ++(0,-0.1);
                \draw ($(#1)+(-0.8,-0.5)$) rectangle ($(#1)+(-1,-0.85)$);
                \draw [/tikz/circuitikz/bipoles/length=0.6cm] ($(#1)+(-0.9,-0.85)$) node[ground]{};
                \draw ($(#1)+(-0.925,-0.7)$) [anchor=east] node{\footnotesize\ensuremath{#3}};
            }
        }
      }

    \ifthenelse{\equal{#4}{delta}}%
      {% Label right side with a delta connection
        \draw ($(#1)+(0.6,-0.45)$) node{\Large $\Delta$};
      }
      {% Label right side with a wye connection
        \draw ($(#1)+(0.6,-0.4)$) to[short, -] ++(-90:0.3) node[left]{};
        \draw ($(#1)+(0.6,-0.4)$) to[short, -] ++(30:0.3)  node[above]{};
        \draw ($(#1)+(0.6,-0.4)$) to[short, -] ++(150:0.3) node[left]{};
        \ifthenelse{\equal{#4}{wye}}%
        {% Draw an ungrounded wye (nothing additional required)
        }%
        {\ifthenelse{\equal{#4}{0}}%
            {% Draw a wye with a neutral shorted to ground
                \draw [/tikz/circuitikz/bipoles/length=0.6cm] ($(#1)+(0.6,-0.4)$) to[short,-] ++(0.3,0) to[short, -] ++(0,-0.1) node[ground]{};
            }
            {% Draw a wye with a neutral grounded through the specified impedance
                \draw [/tikz/circuitikz/bipoles/length=0.6cm] ($(#1)+(0.6,-0.4)$) to[short,-] ++(0.3,0) to[short, -] ++(0,-0.1);
                \draw ($(#1)+(0.8,-0.5)$) rectangle ($(#1)+(1,-0.85)$);
                \draw [/tikz/circuitikz/bipoles/length=0.6cm] ($(#1)+(0.9,-0.85)$) node[ground]{};
                \draw ($(#1)+(0.925,-0.7)$) [anchor=west] node{\footnotesize\ensuremath{#4}};
            }
        }
      }
      
    \draw ($(#1)+(0,0.65)$) [anchor=south] node{\ensuremath{#5}};  

    \ifthenelse{\equal{#3}{delta} \AND \equal{#4}{delta}}%
      {% Both sides are deltas
        \draw ($(#1)+(-0.1,-1)$) [anchor=east] node{\mbox{#6}};
        \draw ($(#1)+(0.1,-1)$) [anchor=west] node{\mbox{#7}};
        \draw ($(#1)+(0,-1.65)$) [anchor=south] node{\mbox{#8}};
        \draw ($(#1)+(0,-2.15)$) [anchor=south] node{\mbox{#9}};
      }
      {\ifthenelse{\equal{#3}{wye} \AND \equal{#4}{wye}}%
        {% Both sides are ungrounded wye
            \draw ($(#1)+(-0.1,-1)$) [anchor=east] node{\mbox{#6}};
            \draw ($(#1)+(0.1,-1)$) [anchor=west] node{\mbox{#7}};
            \draw ($(#1)+(0,-1.65)$) [anchor=south] node{\mbox{#8}};
            \draw ($(#1)+(0,-2.15)$) [anchor=south] node{\mbox{#9}};
        }%
        {\ifthenelse{\equal{#3}{delta} \AND \equal{#4}{wye}}%
            {% Mix of ungrounded wye and delta
                \draw ($(#1)+(-0.1,-1)$) [anchor=east] node{\mbox{#6}};
                \draw ($(#1)+(0.1,-1)$) [anchor=west] node{\mbox{#7}};
                \draw ($(#1)+(0,-1.65)$) [anchor=south] node{\mbox{#8}};
                \draw ($(#1)+(0,-2.15)$) [anchor=south] node{\mbox{#9}};
            }
            {\ifthenelse{\equal{#3}{wye} \AND \equal{#4}{delta}}%
                {% Mix of ungrounded wye and delta
                    \draw ($(#1)+(-0.1,-1)$) [anchor=east] node{\mbox{#6}};
                    \draw ($(#1)+(0.1,-1)$) [anchor=west] node{\mbox{#7}};
                    \draw ($(#1)+(0,-1.65)$) [anchor=south] node{\mbox{#8}};
                    \draw ($(#1)+(0,-2.15)$) [anchor=south] node{\mbox{#9}};
                }
                {\ifthenelse{\equal{#3}{0} \AND \equal{#4}{delta}}%
                    {% Mix of grounded wye and delta
                        \draw ($(#1)+(-0.1,-1)$) [anchor=east] node{\mbox{#6}};
                        \draw ($(#1)+(0.1,-1)$) [anchor=west] node{\mbox{#7}};
                        \draw ($(#1)+(0,-1.65)$) [anchor=south] node{\mbox{#8}};
                        \draw ($(#1)+(0,-2.15)$) [anchor=south] node{\mbox{#9}};
                    }
                    {\ifthenelse{\equal{#3}{delta} \AND \equal{#4}{0}}%
                        {
                            \draw ($(#1)+(-0.1,-1)$) [anchor=east] node{\mbox{#6}};
                            \draw ($(#1)+(0.1,-1)$) [anchor=west] node{\mbox{#7}};
                            \draw ($(#1)+(0,-1.65)$) [anchor=south] node{\mbox{#8}};
                            \draw ($(#1)+(0,-2.15)$) [anchor=south] node{\mbox{#9}};
                        }
                        {\ifthenelse{\equal{#3}{wye} \AND \equal{#4}{0}}%
                            {
                                \draw ($(#1)+(-0.1,-1)$) [anchor=east] node{\mbox{#6}};
                                \draw ($(#1)+(0.1,-1)$) [anchor=west] node{\mbox{#7}};
                                \draw ($(#1)+(0,-1.65)$) [anchor=south] node{\mbox{#8}};
                                \draw ($(#1)+(0,-2.15)$) [anchor=south] node{\mbox{#9}};
                            }
                            {\ifthenelse{\equal{#3}{0} \AND \equal{#4}{wye}}%
                                {
                                    \draw ($(#1)+(-0.1,-1)$) [anchor=east] node{\mbox{#6}};
                                    \draw ($(#1)+(0.1,-1)$) [anchor=west] node{\mbox{#7}};
                                    \draw ($(#1)+(0,-1.65)$) [anchor=south] node{\mbox{#8}};
                                    \draw ($(#1)+(0,-2.15)$) [anchor=south] node{\mbox{#9}};
                                }
                                {\ifthenelse{\equal{#3}{0} \AND \equal{#4}{0}}%
                                    {
                                        \draw ($(#1)+(-0.1,-1)$) [anchor=east] node{\mbox{#6}};
                                        \draw ($(#1)+(0.1,-1)$) [anchor=west] node{\mbox{#7}};
                                        \draw ($(#1)+(0,-1.65)$) [anchor=south] node{\mbox{#8}};
                                        \draw ($(#1)+(0,-2.15)$) [anchor=south] node{\mbox{#9}};
                                    }
                                    {% At least one wye grounded through an impedance
                                        \draw ($(#1)+(-0.1,-1.4)$) [anchor=east] node{\mbox{#6}};
                                        \draw ($(#1)+(0.1,-1.4)$) [anchor=west] node{\mbox{#7}};
                                        \draw ($(#1)+(0,-2.05)$) [anchor=south] node{\mbox{#8}};
                                        \draw ($(#1)+(0,-2.55)$) [anchor=south] node{\mbox{#9}};
                                    }
                                }
                            }
                        }
                    }
                }
            }
        }
      }
}

\newcommand{\generatorL}[7]{%
  \coordinate (#1) at #2;

  \draw (#1) node[vsourcesinshape, rotate=90] (#1_temp) {} ++(0.43,0) to[short,-*] +(0.5,0) coordinate (#1_right);

  \ifthenelse{\equal{#3}{}}%
  {
  }
  {
      \ifthenelse{\equal{#3}{delta}}%
          {% Label with a delta connection
            \draw ($(#1)+(-0.6,-0.55)$) node{\Large $\Delta$};
          }
          {% Label with a wye connection
            \draw ($(#1)+(-0.7,-0.5)$) to[short, -] ++(-90:0.3) node[left]{};
            \draw ($(#1)+(-0.7,-0.5)$) to[short, -] ++(30:0.3)  node[above]{};
            \draw ($(#1)+(-0.7,-0.5)$) to[short, -] ++(150:0.3) node[left]{};
            \ifthenelse{\equal{#3}{wye}}%
            {% Draw an ungrounded wye (nothing additional required)
            }%
            {\ifthenelse{\equal{#3}{0}}%
                {% Draw a wye with a neutral shorted to ground
                    \draw [/tikz/circuitikz/bipoles/length=0.6cm] ($(#1)+(-0.7,-0.5)$) to[short,-] ++(-0.3,0) to[short, -] ++(0,-0.1) node[ground]{};
                }
                {% Draw a wye with a neutral grounded through the specified impedance
                
                    \draw [/tikz/circuitikz/bipoles/length=0.6cm] ($(#1)+(-0.7,-0.5)$) to[short,-] ++(-0.3,0) to[short, -] ++(0,-0.1);
                    \draw ($(#1)+(-1.1,-0.6)$) rectangle ($(#1)+(-0.9,-0.95)$);
                    \draw [/tikz/circuitikz/bipoles/length=0.6cm] ($(#1)+(-1,-0.95)$) node[ground]{};
                    \draw ($(#1)+(-1.1,-0.8)$) [anchor=east] node{\footnotesize #3};
                }
            }
          }
    }
    
    \draw ($(#1)+(-0.4,0)$) [anchor=east] node{\ensuremath{#4}};

    \draw ($(#1)+(-0.4,-0.875)$) [anchor=west,align=left] node{#5};

    \draw (#1_right) [anchor=north] node{\ensuremath{#7}};

    \ifthenelse{\equal{#6}{}}%
    {% No power flow arrow, do nothing
    }
    {
        \path [->,>=stealth] ($(#1)+(0.75,0.3)$) edge ($(#1)+(1.25,0.3)$);
        \draw ($(#1)+(1,0.35)$) [anchor=south] node{\footnotesize #6};
    }
    
}

\newcommand{\generatorR}[7]{%
  \coordinate (#1) at #2;

  \draw (#1) node[vsourcesinshape, rotate=90] (#1_temp) {} ++(-0.43,0) to[short,-*] +(-0.5,0) coordinate (#1_left);

  \ifthenelse{\equal{#3}{}}%
  {
  }
  {
      \ifthenelse{\equal{#3}{delta}}%
          {% Label with a delta connection
            \draw ($(#1)+(0.6,-0.55)$) node{\Large $\Delta$};
          }
          {% Label with a wye connection
            \draw ($(#1)+(0.7,-0.5)$) to[short, -] ++(-90:0.3) node[left]{};
            \draw ($(#1)+(0.7,-0.5)$) to[short, -] ++(30:0.3)  node[above]{};
            \draw ($(#1)+(0.7,-0.5)$) to[short, -] ++(150:0.3) node[left]{};
            \ifthenelse{\equal{#3}{wye}}%
            {% Draw an ungrounded wye (nothing additional required)
            }%
            {\ifthenelse{\equal{#3}{0}}%
                {% Draw a wye with a neutral shorted to ground
                    \draw [/tikz/circuitikz/bipoles/length=0.6cm] ($(#1)+(0.7,-0.5)$) to[short,-] ++(0.3,0) to[short, -] ++(0,-0.1) node[ground]{};
                }
                {% Draw a wye with a neutral grounded through the specified impedance
                
                    \draw [/tikz/circuitikz/bipoles/length=0.6cm] ($(#1)+(0.7,-0.5)$) to[short,-] ++(0.3,0) to[short, -] ++(0,-0.1);
                    \draw ($(#1)+(1.1,-0.6)$) rectangle ($(#1)+(0.9,-0.95)$);
                    \draw [/tikz/circuitikz/bipoles/length=0.6cm] ($(#1)+(1,-0.95)$) node[ground]{};
                    \draw ($(#1)+(1.1,-0.8)$) [anchor=west] node{\footnotesize #3};
                }
            }
          }
    }
    
    \draw ($(#1)+(0.4,0)$) [anchor=west] node{\ensuremath{#4}};

    \draw ($(#1)+(0.4,-0.875)$) [anchor=east,align=left] node{#5};

    \draw (#1_left) [anchor=north] node{\ensuremath{#7}};

    \ifthenelse{\equal{#6}{}}%
    {% No power flow arrow, do nothing
    }
    {
        \path [->,>=stealth] ($(#1)+(-0.75,0.3)$) edge ($(#1)+(-1.25,0.3)$);
        \draw ($(#1)+(-1,0.35)$) [anchor=south] node{\footnotesize #6};
    }
}

\newcommand{\bus}[6]{%
  \coordinate (#1) at #2;

  \path[draw,line width=4pt] ($(#1)+(0,1)$) -- ($(#1)+(0,-1)$);

  \ifthenelse{\equal{#4}{UL}}%
  {%
      \draw ($(#1)+(0,1)$) node[left] {{\raisebox{.5pt}{\textcircled{\raisebox{-.9pt} {#3}}}}};
  }

  \ifthenelse{\equal{#4}{UR}}%
  {%
      \draw ($(#1)+(0,1)$) node[right] {{\raisebox{.5pt}{\textcircled{\raisebox{-.9pt} {#3}}}}};
  }

  \ifthenelse{\equal{#4}{LL}}%
  {%
      \draw ($(#1)+(0,-1)$) node[left] {{\raisebox{.5pt}{\textcircled{\raisebox{-.9pt} {#3}}}}};
  }

  \ifthenelse{\equal{#4}{LR}}%
  {%
      \draw ($(#1)+(0,-1)$) node[right] {{\raisebox{.5pt}{\textcircled{\raisebox{-.9pt} {#3}}}}};
  }

  \ifthenelse{\equal{#5}{}}%
  {% No load demand on the left, do nothing
  }
  {% Draw load demand on the left
      \path[draw,line width=1pt] ($(#1)+(0,-0.8)$) -- ($(#1)+(-0.3,-0.8)$);
      \draw [line width=1pt, ->] ($(#1)+(-0.3,-0.8)$) -- ($(#1)+(-0.3,-1.5)$);
      \ifthenelse{\equal{#5}{0}}%
      {% Load demand = 0 means don't add a label
      }
      {
        \draw ($(#1)+(-0.3,-1.15)$) [anchor=east,align=left] node{#5};
      }
  }

  \ifthenelse{\equal{#6}{}}%
  {% No load demand on the right, do nothing
  }
  {% Draw load demand on the right
      \path[draw,line width=1pt] ($(#1)+(0,-0.8)$) -- ($(#1)+(0.3,-0.8)$);
      \draw [line width=1pt, ->] ($(#1)+(0.3,-0.8)$) -- ($(#1)+(0.3,-1.5)$);
      \ifthenelse{\equal{#6}{0}}%
      {% Load demand = 0 means don't add a label
      }
      {
        \draw ($(#1)+(0.3,-1.15)$) [anchor=west,align=left] node{#6};
      }
  }

  \coordinate (#1_2_2) at ($(#1)+(0,-0.5)$);
  \coordinate (#1_2_1) at ($(#1)+(0,0.5)$);

  \coordinate (#1_3_3) at ($(#1)+(0,-0.6)$);
  \coordinate (#1_3_2) at ($(#1)+(0,0)$);
  \coordinate (#1_3_1) at ($(#1)+(0,0.6)$);

  \coordinate (#1_4_4) at ($(#1)+(0,-0.6)$);
  \coordinate (#1_4_3) at ($(#1)+(0,-0.2)$);
  \coordinate (#1_4_2) at ($(#1)+(0,0.2)$);
  \coordinate (#1_4_1) at ($(#1)+(0,0.6)$);

  \coordinate (#1_5_5) at ($(#1)+(0,-0.8)$);
  \coordinate (#1_5_4) at ($(#1)+(0,-0.4)$);
  \coordinate (#1_5_3) at ($(#1)+(0,0)$);
  \coordinate (#1_5_2) at ($(#1)+(0,0.4)$);
  \coordinate (#1_5_1) at ($(#1)+(0,0.8)$);

}

\newcommand{\straightline}[7]{%
  \coordinate (#1_1) at (#2);
  \coordinate (#1_2) at (#3);

  \path[draw,line width=2pt] (#1_1) -- (#1_2);
  
  \node[anchor=west, align=left] (#1_midpt) at ($(#1_1)!0.5!(#1_2) + #5$) {#4};

  \ifthenelse{\equal{#6}{}}%
  {% No first breaker specified, do nothing
  }
  {% Add first breaker
    \coordinate (#1_B1_loc) at ($(#1_1)!0.15!(#1_2)$);
    \draw [fill=black,draw=none] ($(#1_B1_loc)+(0.125,0.125)$) rectangle ($(#1_B1_loc)+(-0.125,-0.125)$);
    \ifthenelse{\equal{#6}{0}}%
    {% Don't add label
    }
    {% Add label
      \draw ($(#1_B1_loc)+(0,0.1)$) [anchor=south] node{\footnotesize #6};
    }
  }

  \ifthenelse{\equal{#7}{}}%
  {% No second breaker specified, do nothing
  }
  {% Add second breaker
    \coordinate (#1_B2_loc) at ($(#1_1)!0.85!(#1_2)$);
    \draw [fill=black,draw=none] ($(#1_B2_loc)+(0.125,0.125)$) rectangle ($(#1_B2_loc)+(-0.125,-0.125)$);
    \ifthenelse{\equal{#7}{0}}%
    {% Don't add label
    }
    {% Add label
      \draw ($(#1_B2_loc)+(0,0.1)$) [anchor=south] node{\footnotesize #7};
    }
  }
}

\newcommand{\Cline}[9]{%
  \coordinate (#1_1) at (#2);
  \coordinate (#1_2) at (#3);

  \path[draw,line width=2pt] (#1_1) -- ++(#4,#5) coordinate (#1_temp1);
  \path[draw,line width=2pt] (#1_2) -- ++(#4,-#5) coordinate (#1_temp2);
  \path[draw,line width=2pt] (#1_temp1) -- (#1_temp2);
  
  \node[anchor=west, align=left] (#1_midpt) at ($(#1_temp1)!0.5!(#1_temp2) + #7$) {#6};

  \ifthenelse{\equal{#8}{}}%
  {% No first breaker specified, do nothing
  }
  {% Add first breaker
    \coordinate (#1_B1_loc) at ($(#1_1)+(#4/2,#5/2)$);
    \draw [fill=black,draw=none] ($(#1_B1_loc)+(0.125,0.125)$) rectangle ($(#1_B1_loc)+(-0.125,-0.125)$);
    \ifthenelse{\equal{#8}{0}}%
    {% Don't add label
    }
    {% Add label
      \draw ($(#1_B1_loc)+(0,0.1)$) [anchor=south] node{\footnotesize #8};
    }
  }

  \ifthenelse{\equal{#9}{}}%
  {% No second breaker specified, do nothing
  }
  {% Add second breaker
    \coordinate (#1_B2_loc) at ($(#1_2)+(#4/2,-#5/2)$);
    \draw [fill=black,draw=none] ($(#1_B2_loc)+(0.125,0.125)$) rectangle ($(#1_B2_loc)+(-0.125,-0.125)$);
    \ifthenelse{\equal{#9}{0}}%
    {% Don't add label
    }
    {% Add label
      \draw ($(#1_B2_loc)+(0,0.1)$) [anchor=south] node{\footnotesize #9};
    }
  }
}

\newcommand{\Sline}[9]{%
  \coordinate (#1_1) at (#2);
  \coordinate (#1_2) at (#3);

  \path[draw,line width=2pt] (#1_1) -- ++(#4,#5) coordinate (#1_temp1);
  \path[draw,line width=2pt] (#1_2) -- ++(-#4,-#5) coordinate (#1_temp2);
  \path[draw,line width=2pt] (#1_temp1) -- (#1_temp2);
  
  \node[anchor=west, align=left] (#1_midpt) at ($(#1_temp1)!0.5!(#1_temp2) + #7$) {#6};

  \ifthenelse{\equal{#8}{}}%
  {% No first breaker specified, do nothing
  }
  {% Add first breaker
    \ifthenelse{\lengthtest{#4pt < 1.5pt}}% Comparing floating point numbers in Latex seems surprisingly hard. This isn't ideal but should cover most of the relevant cases.
    {% Cap the location of the breaker to be close to the bus
        \coordinate (#1_B1_loc) at ($(#1_1)+(#4/2,#5/2)$);
    }
    {
        \coordinate (#1_B1_loc) at ($(#1_1)+({#4/sqrt(#4^2+#5^2)/2},{#5/sqrt(#4^2+#5^2)/2})$);
    }
    \draw [fill=black,draw=none] ($(#1_B1_loc)+(0.125,0.125)$) rectangle ($(#1_B1_loc)+(-0.125,-0.125)$);
    \ifthenelse{\equal{#8}{0}}%
    {% Don't add label
    }
    {% Add label
      \draw ($(#1_B1_loc)+(0,0.1)$) [anchor=south] node{\footnotesize #8};
    }
  }

  \ifthenelse{\equal{#9}{}}%
  {% No second breaker specified, do nothing
  }
  {% Add second breaker
      \ifthenelse{\lengthtest{#4pt < 1.5pt}}% Comparing floating point numbers in Latex seems surprisingly hard. This isn't ideal but should cover most of the relevant cases.
    {% Cap the location of the breaker to be close to the bus
        \coordinate (#1_B2_loc) at ($(#1_2)+(-#4/2,-#5/2)$);
    }
    {
        \coordinate (#1_B2_loc) at ($(#1_2)+({-#4/sqrt(#4^2+#5^2)/2},{-#5/sqrt(#4^2+#5^2)/2})$);
    }
    \draw [fill=black,draw=none] ($(#1_B2_loc)+(0.125,0.125)$) rectangle ($(#1_B2_loc)+(-0.125,-0.125)$);
    \ifthenelse{\equal{#9}{0}}%
    {% Don't add label
    }
    {% Add label
      \draw ($(#1_B2_loc)+(0,0.1)$) [anchor=south] node{\footnotesize #9};
    }
  }
}

\newcommand{\wyedeltaequivalent}[5]{%
  \coordinate (#1) at #2;
  \coordinate (#1_left) at (#1);

  \draw (#1) to[R=\mbox{#3},-] ++(2,0) coordinate (#1_shift1p) to[open] ++(0.5,0) -- ++(0.5,0) coordinate (#1_right);

  \draw ($(#1_shift1p)+(0.25,0)$) circle (0.25cm);
  \ifthenelse{\equal{#5}{left}}%
  {
    \draw ($(#1_shift1p)+(0,-0.2)$) -- ($(#1_shift1p)+(0.5,0.2)$);
  }
  {
    \draw ($(#1_shift1p)+(0,0.2)$) -- ($(#1_shift1p)+(0.5,-0.2)$);
  }
  
  \node (#1_shift1p_label) at ($(#1_shift1p)+(0.25,-0.55)$) {\mbox{\footnotesize\ensuremath{#4}}};
}

\newcommand{\switchedCapacitor}[4]{%
  \coordinate (#1) at (#2);
  \draw (#1) to[C=#3,*-*] +({0+90}:1.25);
  \draw ($(#1)+(0,3)$) to[short,*-] ($(#1)+(0,2.5)$);
  \draw ($(#1)+(0,2.5)$) -- ++(0,-0.2) 
        node (#1_temp) [draw, inner sep=1pt, circle, fill=white]{} 
        ++(0,-0.6) node [draw,inner sep=1pt, circle, fill=white]{} 
        -- ++(0,-0.2);
    \draw [thick] (#1_temp) -- ++(0.3, -0.5);
    \draw (#1_temp) ++(0,-0.6) to[short,-] ($(#1)+({0+90}:1.25)$);
  \node (#1_label) at ($(#1)+(-0.50,2)$){#4};
}

\newcommand{\waveformplotSixty}[8]{%

    \providecommand\waveformlabel{}
    \providecommand\waveformunits{}
    \providecommand\labelshift{}
    \providecommand\ytick{}
    \providecommand\xtick{}
    \providecommand\xtickLabel{}
    
    \pgfmathsetmacro{\waveformIndex}{#1} % Location of the waveform on the page
    \pgfmathsetmacro{\amp}{#2} % Amplitude of the waveform
    \pgfmathsetmacro{\theta}{#3} % Phase of the waveform in the form \amp*cos(omega*t + \theta)
    \renewcommand{\waveformlabel}[0]{#4} % Label for the waveform at the top of the plot
    \renewcommand{\waveformunits}[0]{#5} % Units for the waveform (typically, V, A, or W)
    \renewcommand{\labelshift}[0]{#6} % Controls where on the plot the amplitude labels will show up
    \pgfmathsetmacro{\xtickOpt}{#7} % 30 or 45 degree increments
    \pgfmathsetmacro{\ytickOpt}{#8} % 1 or 2 unit increments
    
    \ifthenelse{\equal{\numPlots}{4}}%
    {
        \pgfmathsetmacro{\plotHeight}{2.3in}
        \ifthenelse{\equal{\waveformIndex}{1}}%
        {
            \node (label) at (0.5,6) {\waveformlabel{}~[\waveformunits{}]};
            \pgfmathsetmacro{\yshift}{1.25cm}
        }
        {
            \ifthenelse{\equal{\waveformIndex}{2}}%
            {
                \node (label) at (0.5,0.5) {\waveformlabel{}~[\waveformunits{}]};
                \pgfmathsetmacro{\yshift}{-4.25cm}
            }
            {
                \ifthenelse{\equal{\waveformIndex}{3}}%
                {
                    \node (label) at (0.5,-5) {\waveformlabel{}~[\waveformunits{}]};
                    \pgfmathsetmacro{\yshift}{-9.75cm}
                }
                {
                    \ifthenelse{\equal{\waveformIndex}{4}}%
                    {
                        \node (label) at (0.5,-10.5) {\waveformlabel{}~[\waveformunits{}]};
                        \pgfmathsetmacro{\yshift}{-15.25cm}
                    }
                    {
                    }
                }
            }
        }
    }
    {% If there are three plots on the page
        \pgfmathsetmacro{\plotHeight}{3in}
        \ifthenelse{\equal{\waveformIndex}{1}}%
        {
            \node (label) at (0.5,7.75) {\waveformlabel{}~[\waveformunits{}]};
            \pgfmathsetmacro{\yshift}{1.25cm}
        }
        {
            \ifthenelse{\equal{\waveformIndex}{2}}%
            {
                \node (label) at (0.5,0.5) {\waveformlabel{}~[\waveformunits{}]};
                \pgfmathsetmacro{\yshift}{-6cm}
            }
            {
                \ifthenelse{\equal{\waveformIndex}{3}}%
                {
                    \node (label) at (0.5,-6.75) {\waveformlabel{}~[\waveformunits{}]};
                    \pgfmathsetmacro{\yshift}{-13.25cm}
                }
                {
                }
            }
        }
    }
    \ifthenelse{\equal{\xtickOpt}{30}}%
    {
        \renewcommand{\xtick}[0]{0,30,60,90,120,150,180,210,240,270,300,330,360,390,420,450,480}
    }
    {
        \ifthenelse{\equal{\xtickOpt}{45}}%
        {
            \renewcommand{\xtick}[0]{0,45,90,135,180,225,270,315,360,405,450}
        }
        {
        }
    }
    
    \ifthenelse{\equal{\ytickOpt}{1}}%
    {
        \renewcommand{\ytick}[0]{-6,-5,-4,-3,-2,-1,0,1,2,3,4,5,6}
        \pgfmathsetmacro{\ytickMin}{-6}
        \pgfmathsetmacro{\ytickMax}{6}
    }
    {
        \ifthenelse{\equal{\ytickOpt}{2}}%
        {
            \renewcommand{\ytick}[0]{-10,-8,-6,-4,-2,0,2,4,6,8,10}
            \pgfmathsetmacro{\ytickMin}{-10}
            \pgfmathsetmacro{\ytickMax}{10}
        }
        {
            \ifthenelse{\equal{\ytickOpt}{5}}%
            {
                \renewcommand{\ytick}[0]{-25,-20,-15,-10,-5,0,5,10,15,20,25}
                \pgfmathsetmacro{\ytickMin}{-25}
                \pgfmathsetmacro{\ytickMax}{25}
            }
            {
                \ifthenelse{\equal{\ytickOpt}{10}}%
                {
                    \renewcommand{\ytick}[0]{-50,-40,-30,-20,-10,0,10,20,30,40,50}
                    \pgfmathsetmacro{\ytickMin}{-50}
                    \pgfmathsetmacro{\ytickMax}{50}
                }
                {
                }
            }
        }
    }
    
    \pgfmathsetmacro{\xMax}{480}
    \pgfmathsetmacro{\xMin}{-10}
    
    \begin{axis}[width=\textwidth,
        height=\plotHeight,
        axis y line=middle,
        axis x line=middle,
        xlabel = $\omega t$,
        ylabel = {},
        xtick=\xtick,
        xticklabel={$\pgfmathprintnumber{\tick}^\circ$},
        ytick=\ytick, ymin=\ytickMin, ymax=\ytickMax,
        yshift=\yshift,
        clip=false
        ]
    \addplot [domain=\xMin:\xMax, 
        samples=100]{\amp*cos(x+\theta)};
    \addplot [dotted,
        domain=0:\xMax, 
        samples=100]{\amp};
    \addplot [dotted,
        domain=0:\xMax, 
        samples=100]{-\amp};

    \pgfmathparse{-\theta >= \xMin ? 1 : 0}
    \ifthenelse {\pgfmathresult>0}
    {
        \addplot [dotted] coordinates {
        (0-\theta, \amp)
        (0-\theta, 0)};    
    }
    { % Don't plot zeros outside the range
    }
    \addplot [dotted] coordinates {
            (180-\theta, -\amp)
            (180-\theta, 0)};
    \pgfmathparse{-\theta+360 <= \xMax ? 1 : 0}
    \ifthenelse {\pgfmathresult>0}
    {
        \addplot [dotted] coordinates {
        (360-\theta, \amp)
        (360-\theta, 0)};
    }
    { % Don't plot zeros outside the range
    }
        
    \ifthenelse{\equal{\amp}{0}}%
    {% Amplitude is zero, so don't plot amplitude labels
    }
    {
        \ifthenelse{\equal{\ampOrientation}{2}}%
        {
            \node[label={270:{\amp{}~\waveformunits{}}},yshift=3] at (axis cs:\labelshift,\amp) {};
            \node[label={90:{-\amp{}~\waveformunits{}}},yshift=-3] at (axis cs:\labelshift,-\amp) {};
        }
        {
            \node[label={90:{\amp{}~\waveformunits{}}},yshift=-3] at (axis cs:\labelshift,\amp) {};
            \node[label={270:{-\amp{}~\waveformunits{}}},yshift=3] at (axis cs:\labelshift,-\amp) {};
        }

        \pgfmathparse{-\theta+90-180 >= \xMin ? 1 : 0}
        \ifthenelse {\pgfmathresult>0}
        {
            \node[circle,fill,inner sep=1pt] at (axis cs:-\theta+90-180,0) {};
        }
        \pgfmathparse{-\theta+90 >= \xMin ? 1 : 0}
        \ifthenelse {\pgfmathresult>0}
        {
            \node[circle,fill,inner sep=1pt] at (axis cs:-\theta+90,0) {};
        }
        { % Don't plot zeros outside the range
        }
        \node[circle,fill,inner sep=1pt] at (axis cs:-\theta+90+180,0) {};
        \pgfmathparse{-\theta+90+360 <= \xMax ? 1 : 0}
        \ifthenelse {\pgfmathresult>0}
        {
            \node[circle,fill,inner sep=1pt] at (axis cs:-\theta+90+360,0) {};
        }
        { % Don't plot zeros outside the range
        } 
    }
    \end{axis}
}

\newcommand{\waveformplotOneTwenty}[9]{%

    \providecommand\waveformlabel{}
    \providecommand\waveformunits{}
    \providecommand\labelshift{}
    \providecommand\ytick{}
    \providecommand\xtick{}
    \providecommand\xtickLabel{}
    
    \pgfmathsetmacro{\waveformIndex}{#1} % Location of the waveform on the page
    \pgfmathsetmacro{\amp}{#2} % Amplitude of the waveform
    \pgfmathsetmacro{\theta}{#3} % Phase of the first waveform in degrees
    \pgfmathsetmacro{\phi}{#4} % Phase of the second waveform in degrees
    \renewcommand{\waveformlabel}[0]{#5} % Label for the waveform at the top of the plot
    \renewcommand{\waveformunits}[0]{#6} % Units for the waveform (typically, V, A, or W)
    \renewcommand{\labelshift}[0]{#7} % Controls where on the plot the amplitude labels will show up
    \pgfmathsetmacro{\xtickOpt}{#8} % 30 or 45 degree increments
    \pgfmathsetmacro{\ytickOpt}{#9} % 1, 2, 5, or 10 unit increments

    \pgfmathsetmacro{\xMax}{480}
    \pgfmathsetmacro{\xMin}{-10}

    \pgfmathsetmacro{\waveformMax}{0.5*\amp*(cos(\theta-\phi)+1)} % This is the actual maximum of the double-frequency waveform
    \pgfmathsetmacro{\waveformMin}{0.5*\amp*(cos(\theta-\phi)-1)} % This is the actual minimum of the double-frequency waveform
    \pgfmathsetmacro{\waveformAvg}{0.5*(\waveformMax+\waveformMin)} % This is the actual minimum of the double-frequency waveform
    
    \ifthenelse{\equal{\numPlots}{4}}%
    {
        \pgfmathsetmacro{\plotHeight}{2.3in}
        \ifthenelse{\equal{\waveformIndex}{1}}%
        {
            \node (label) at (0.5,6) {\waveformlabel{}~[\waveformunits{}]};
            \pgfmathsetmacro{\yshift}{1.25cm}
        }
        {
            \ifthenelse{\equal{\waveformIndex}{2}}%
            {
                \node (label) at (0.5,0.5) {\waveformlabel{}~[\waveformunits{}]};
                \pgfmathsetmacro{\yshift}{-4.25cm}
            }
            {
                \ifthenelse{\equal{\waveformIndex}{3}}%
                {
                    \node (label) at (0.5,-5) {\waveformlabel{}~[\waveformunits{}]};
                    \pgfmathsetmacro{\yshift}{-9.75cm}
                }
                {
                    \ifthenelse{\equal{\waveformIndex}{4}}%
                    {
                        \node (label) at (0.5,-10.5) {\waveformlabel{}~[\waveformunits{}]};
                        \pgfmathsetmacro{\yshift}{-15.25cm}
                    }
                    {
                    }
                }
            }
        }
    }
    {% If there are three plots on the page
        \pgfmathsetmacro{\plotHeight}{3in}
        \ifthenelse{\equal{\waveformIndex}{1}}%
        {
            \node (label) at (0.5,7.75) {\waveformlabel{}~[\waveformunits{}]};
            \pgfmathsetmacro{\yshift}{1.25cm}
        }
        {
            \ifthenelse{\equal{\waveformIndex}{2}}%
            {
                \node (label) at (0.5,0.5) {\waveformlabel{}~[\waveformunits{}]};
                \pgfmathsetmacro{\yshift}{-6cm}
            }
            {
                \ifthenelse{\equal{\waveformIndex}{3}}%
                {
                    \node (label) at (0.5,-6.75) {\waveformlabel{}~[\waveformunits{}]};
                    \pgfmathsetmacro{\yshift}{-13.25cm}
                }
                {
                }
            }
        }
    }
    \ifthenelse{\equal{\xtickOpt}{30}}%
    {
        \renewcommand{\xtick}[0]{0,30,60,90,120,150,180,210,240,270,300,330,360,390,420,450}
    }
    {
        \ifthenelse{\equal{\xtickOpt}{45}}%
        {
            \renewcommand{\xtick}[0]{0,30,60,90,120,150,180,210,240,270,300,330,360,390,420,450}
        }
        {
        }
    }
    
    \ifthenelse{\equal{\ytickOpt}{1}}%
    {
        \renewcommand{\ytick}[0]{-6,-5,-4,-3,-2,-1,0,1,2,3,4,5,6}
        \pgfmathsetmacro{\ytickMin}{-6}
        \pgfmathsetmacro{\ytickMax}{6}
    }
    {
        \ifthenelse{\equal{\ytickOpt}{2}}%
        {
            \renewcommand{\ytick}[0]{-10,-8,-6,-4,-2,0,2,4,6,8,10}
            \pgfmathsetmacro{\ytickMin}{-10}
            \pgfmathsetmacro{\ytickMax}{10}
        }
        {
            \ifthenelse{\equal{\ytickOpt}{5}}%
            {
                \renewcommand{\ytick}[0]{-25,-20,-15,-10,-5,0,5,10,15,20,25}
                \pgfmathsetmacro{\ytickMin}{-25}
                \pgfmathsetmacro{\ytickMax}{25}
            }
            {
                \ifthenelse{\equal{\ytickOpt}{10}}%
                {
                    \renewcommand{\ytick}[0]{-50,-40,-30,-20,-10,0,10,20,30,40,50}
                    \pgfmathsetmacro{\ytickMin}{-50}
                    \pgfmathsetmacro{\ytickMax}{50}
                }
                {
                }
            }
        }
    }
    
    \begin{axis}[width=\textwidth,
        height=\plotHeight,
        axis y line=middle,
        axis x line=middle,
        xlabel = $\omega t$,
        ylabel = {},
        xtick=\xtick,
        xticklabel={$\pgfmathprintnumber{\tick}^\circ$},
        ytick=\ytick, ymin=\ytickMin, ymax=\ytickMax,
        yshift=\yshift,
        clip=false
        ]
    \addplot [domain=\xMin:\xMax, 
        samples=100]{\amp*cos(x+\theta)*cos(x+\phi)};
    \addplot [dotted,
        domain=0:\xMax, 
        samples=100]{\waveformMax};
    \addplot [dotted,
        domain=0:\xMax, 
        samples=100]{\waveformMin};

    \ifthenelse{\equal{\amp}{0}}%
    {% Amplitude is zero, so don't plot amplitude labels
    }
    {
        \ifthenelse{\equal{\ampOrientation}{2}}%
        {
            \node[label={270:{\waveformMax{}~\waveformunits{}}},yshift=3] at (axis cs:\labelshift,\waveformMax) {};
            \node[label={90:{\waveformMin{}~\waveformunits{}}},yshift=-3] at (axis cs:\labelshift,\waveformMin) {};
        }
        {
            \node[label={90:{\waveformMax{}~\waveformunits{}}},yshift=-3] at (axis cs:\labelshift,\waveformMax) {};
            \node[label={270:{\waveformMin{}~\waveformunits{}}},yshift=3] at (axis cs:\labelshift,\waveformMin) {};
        }
        \pgfmathparse{-\theta+90 >= \xMin ? 1 : 0}
        \ifthenelse {\pgfmathresult>0}
        {
            \node[circle,fill,inner sep=1pt] at (axis cs:90-\theta,0) {};
        }
        { % Don't plot zeros outside the range
        }
        
        \pgfmathparse{-\phi+90 >= \xMin ? 1 : 0}
        \ifthenelse {\pgfmathresult>0}
        {
            \node[circle,fill,inner sep=1pt] at (axis cs:90-\phi,0) {};
        }
        { % Don't plot zeros outside the range
        }
        
        \pgfmathparse{-\theta+270 <= \xMax ? 1 : 0}
        \ifthenelse {\pgfmathresult>0}
        {
            \node[circle,fill,inner sep=1pt] at (axis cs:270-\theta,0) {};
        }
        { % Don't plot zeros outside the range
        }

        \pgfmathparse{-\phi+270 <= \xMax ? 1 : 0}
        \ifthenelse {\pgfmathresult>0}
        {
            \node[circle,fill,inner sep=1pt] at (axis cs:270-\phi,0) {};
        }
        { % Don't plot zeros outside the range
        }

        \pgfmathparse{-\theta+450 <= \xMax ? 1 : 0}
        \ifthenelse {\pgfmathresult>0}
        {
            \node[circle,fill,inner sep=1pt] at (axis cs:450-\theta,0) {};
        }
        { % Don't plot zeros outside the range
        }

        \pgfmathparse{-\phi+450 <= \xMax ? 1 : 0}
        \ifthenelse {\pgfmathresult>0}
        {
            \node[circle,fill,inner sep=1pt] at (axis cs:450-\phi,0) {};
        }
        { % Don't plot zeros outside the range
        }
        
        \node[label={90:{\waveformAvg{}~\waveformunits{}}},yshift=-3] at (axis cs:\labelshift+30,\waveformAvg) {};
        \addplot [dashed,
            domain=0:\xMax, 
            samples=100]{0.5*(\waveformMax+\waveformMin)};
    }
    \end{axis}
}

\title{VArsity: Can Large Language Models Keep\\ Power Engineering Students in Phase?}
\author{Samuel Talkington and Daniel K. Molzahn
\thanks{Support from NSF AI Institute for Advances in Optimization (AI4OPT) \#2112533. The work of S.~Talkington is supported by the National Science Foundation Graduate Research Fellowship Program under Grant No. DGE-1650044. Any opinions, findings, and conclusions or recommendations expressed in this material are those of the author(s) and do not necessarily reflect the views of the National Science Foundation.
}%
\thanks{
The authors are with the School of Electrical and Computer Engineering, Georgia Institute of Technology. Email: \{talkington,molzahn\}@gatech.edu.
}%
}
\date{January 2025}

\begin{document}

\maketitle

\begin{abstract}
    This paper provides an educational case study regarding our experience in deploying ChatGPT Large Language Models (LLMs) in the Spring 2025 and Fall 2023 offerings of ECE 4320: Power System Analysis \& Control at Georgia Tech. As part of course assessments, students were tasked with identifying, explaining, and correcting errors in the ChatGPT outputs corresponding to power factor correction problems. While most students successfully identified the errors in the outputs from the \texttt{GPT-4} version of ChatGPT used in Fall 2023, students found the errors from the ChatGPT \texttt{o1} version much more difficult to identify in Spring 2025. As shown in this case study, the role of LLMs in pedagogy, assessment, and learning in power engineering classrooms is an important topic deserving further investigation.
\end{abstract}

\section{Introduction}
In just a few short years, large language models (LLMs) have been ubiquitously deployed in a broad array of fields, including power engineering~\cite{majumder_exploring_2024}. The impacts of these models have been rapid and widespread, especially in higher education~\cite{milano_large_2023,kasneci_chatgpt_2023,wang_examining_2024}. From engineering \cite{wang_examining_2024}, operations research \cite{klamkin2025dualschoolreliablellmsoptimization}, and algorithms~\cite{ding_proof_based_2025}, academic communities are asking critical questions about the role of LLMs. These questions are also being raised in power engineering education~\cite{hickey_llm_power_edu_2024}.

% While LLMs are now everyday tools across engineering workflows, 
The power engineering community has had limited discussion on the impacts of LLMs. Early studies report both promise and peril: while LLMs accelerate ideation, they also confidently generate physically inconsistent or numerically incorrect answers~\cite{milano_large_2023,kasneci_chatgpt_2023}. Electric power systems, as critical infrastructure, may be particularly vulnerable to the risks posed by LLMs, and these risks are especially poignant in the context of power engineering education. If the next generation of power engineers is trained in a way that validates or even promotes frequent errors in units, sign conventions, or phasor arithmetic, this can propagate directly into megawatt-scale mistakes. 

Recognizing that LLMs are likely here to stay in the classroom \cite{milano_large_2023}, there is an urgent need for community discussion on governance and best practices for their use. Prior work emphasizes that teachers and learners must develop new competencies (e.g., prompt strategies, fact checking) to harness LLMs responsibly \cite{kasneci_chatgpt_2023,wang_examining_2024}. In this spirit, prompt design and assessment techniques that integrate LLMs into coursework in a way that reinforces fundamental engineering skills, rather than short-circuiting them, are poised to become essential. 
%This need is especially pressing, as instructors themselves turn to LLMs to save time in generating course content, assisting with grading, or even drafting solution keys~\cite{hill_professors_2025}.

This issue became salient to the authors during our experience teaching the Spring 2025 and Fall 2023 iterations of \emph{ECE 4320: Power System Analysis \& Control},
%\footnote{See \url{https://ece.gatech.edu/courses/ece4320} for the ECE 4320 curriculum.} 
a first course in power systems at Georgia Tech that is typically taken by senior undergraduate and first-year graduate students. In this course, we experimented with deploying OpenAI's ChatGPT LLM as part of exam assessments. Specifically, students were tasked with interpreting and correcting errors in the solutions produced by ChatGPT to various power systems problems.

Using power factor correction problems as a case study topic, the contributions of this paper are:
\begin{enumerate}
    \item \textit{Assessment protocol}: Case study examples using LLM outputs in power engineering course assessments, including a grading rubric based on error identification and correction.
   \item \textit{Empirical data}: A retrospective analysis student success in identifying and correcting LLM output errors.
   \item \textit{Perspectives and guidelines:} Our findings in Section~\ref{sec:results} suggest an opportunity for power engineering educators to refine their assessment practices by harnessing
   LLMs to promote critical engineering skills.
\end{enumerate}

The remainder of this paper is organized as follows. Section~\ref{sec:background}
surveys related work and the ECE 4320 course. Section~\ref{sec:power_factor_problem} details the problem and prompt design. Section~\ref{sec:solution} explains the solution output, the errors it contains, and corrections. Section~\ref{sec:results} presents empirical results. 
% Section~\ref{sec:discussion}
% discusses implications;  
Section~\ref{sec:conclusion} concludes the paper.

\vspace{-1em}
\section{Background}
\label{sec:background}

\subsection{Related work}
\label{sec:related}

Reference~\cite{majumder_exploring_2024} maps research questions for LLMs in the electric energy sector, including operation and planning. Although~\cite{majumder_exploring_2024} did not investigate domain-specific educational impacts, the work of~\cite{hickey_llm_power_edu_2024} evaluated GPT-3.5 on economic dispatch homework problems, finding that 58\% of its numerical outputs violated basic power balance constraints. Two key gaps in prior literature are:
\begin{enumerate}
    \item \emph{Problem setting}: The prior education task explored in \cite{hickey_llm_power_edu_2024}, while valuable, was a short, single-step economic dispatch problem. The economic dispatch problem, including the formulation in \cite{hickey_llm_power_edu_2024}, typically uses a highly simplified model that does not encode the physics that governs the underlying power system. Presently, the literature lacks an understanding of how LLMs understand multielement linear circuits in power engineering assignments.
  \item \emph{Students in the loop}: We currently know little about whether students can identify and correct LLM mistakes in electrical engineering contexts, and which scaffolds help. As LLM workflows become increasingly intertwined in classrooms and engineering practice, it is essential to improve our understanding of students' LLM interactions.
\end{enumerate}
In adjacent communities, \cite{klamkin2025dualschoolreliablellmsoptimization} explores the use of LLMs in optimization education, covering the computation of dual linear programs. Reference~\cite{ding_proof_based_2025} covers the performance of ChatGPT \texttt{o1} for a proof-based analysis of algorithms course, finding that it performed at a passing undergraduate level.

\subsection{Overview of the course and the Georgia Tech experience}
\label{sec:setup}
At Georgia Tech, ECE 4320 is typically one of the final courses taken by undergraduate ECE students who choose the \emph{Electric Energy Systems Thread}, which is a structured curriculum for students who wish to specialize in electric power systems or power electronics. For many such students, the ECE 4320 classroom is the last time they will be exposed to formal power system concepts in classroom settings. 

This course introduces students to key power engineering concepts, such as the power flow equations, the optimal power flow problem, unit commitment, and state estimation. The course thus lays the foundation for a career in power engineering and introduces core concepts for graduate education and research in power systems. The foundational nature of this course means that understanding LLMs' impact is essential.

\section{A Case Study: Power Factor Correction}
\label{sec:power_factor_problem}
In the Spring 2025 iteration of ECE 4320, both students and ChatGPT version \texttt{o1} were presented with a power factor correction problem based on the single-phase, 60~Hz circuit in Fig.~\ref{fig:circuit_for_the_pf_correction_problem}. We focus on this problem as a case study on LLMs in power engineering education. Providing a comparison to an earlier version of ChatGPT (version \texttt{GPT-4}), the Fall 2023 iteration of ECE 4320 posed related power factor correction and AC power flow formulation problems described in Appendices \ref{apdx:gpt-4-pf-correction} and \ref{apdx:gpt4-ac-pf}, respectively. ChatGPT version \texttt{o3} has vastly improved performance on all problems; see Appendix \ref{apdx:new_models}.
%In addition, the Fall 2023 iteration of ECE 4320 also posed an AC power flow formulation problem, described in Appendix \ref{apdx:gpt4-ac-pf}, which did not appear in Spring 2025.  Lastly, as shown in Appendix \ref{apdx:new_models}, the recently released ChatGPT version \texttt{o3} successfully solved all the problems mentioned above.

\subsection{Problem formulation}
\label{sec:prob-form}
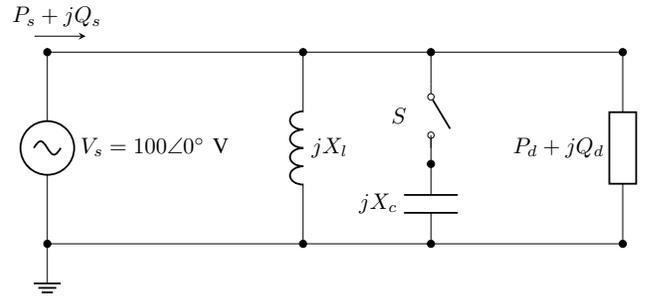
\begin{figure}[t]
\centering
\scalebox{0.85}{
\begin{circuitikz}[american currents,european resistors]
  %% Source
  \coordinate (Csource);
  \draw +({90}:3) coordinate (Sa) node[anchor={0}]{} to[vsourcesin,l^=\mbox{$V_s = 100\angle 0^\circ$~V},*-*] (Csource);
  
  \node (Sg) at ($(Csource)+(0.15,3.55)$){$P_s + j Q_s$};
  \path [->,>=stealth] ($(Csource)+(-0.2,3.25)$) edge ($(Csource)+(0.6,3.25)$);
  
  %% Loads
  \coordinate (Cload1) at ($(Csource)+(9,0)$);
  \draw (Cload1) to[R,l^=\mbox{$P_d + jQ_d$},*-*] +({0+90}:3) coordinate (La) node[anchor=south]{};

  %% Fixed inductor
  \coordinate (FC1) at ($(Csource)+(4,0)$);
  \draw (FC1) to[L,l_=$jX_{l}$,*-*] +({0+90}:3) node[anchor=south]{};

  %% Switched Capacitor
  \switchedCapacitor{SC1}{$(Csource)+(6,0)$}{$jX_{c}$}{$S$}
 
  % %% Connections
  \draw (Sa) -- +(0,0) to[short] ($(La)+(0,0)$) -- (La);
  \draw (0,0) -- ($(La)+(0,-3)$);
  
  \draw ($(Csource)+(0,0)$) -- +(0,-0.2) node[ground]{};

\end{circuitikz}}
\caption{The circuit presented to ChatGPT \texttt{o1} and ECE 4320 students as the context for the power factor correction problem.}
\label{fig:circuit_for_the_pf_correction_problem}
\vspace{-1em}
\end{figure}

In Fig.~\ref{fig:circuit_for_the_pf_correction_problem}, the active power consumed by the load, $P_d$, is a constant value of $50$~W. The reactive power consumed by the load, $Q_d$, continuously varies between a minimum value of ${-}36$~VAr and a maximum value of $60$~VAr. As shown in the circuit above, the voltage source has an RMS value of $100$~V.

The students were instructed to act as circuit designers\textemdash they could choose (fixed) values for both capacitive reactance $X_l$ and inductive reactance $X_c$, as well as the switching configurations for $S$ as a function of the changing reactive power demand of the load. Within this problem setting, payments to the utility company that owns the voltage source $V_s$ were modeled as being based on the smallest power factor of the power $P_s + jQ_s$ supplied across the range of reactive demands $Q_d$. 

Thus, the goal of the problem statement was to choose $X_l$, $X_c$, and a strategy for configuring the switch $S$ to make the smallest power factor of $P_s + jQ_s$ as large as possible (that is, as close to 1.0, either lagging or leading) as the load’s reactive power demand $Q_d$ varies within the range from ${-}36$~Var (supplying $36$~VAr) to $60$~VAr (consuming $60$~VAr).

\subsection{Power factor correction prompt}
\label{sec:prompt}

The ChatGPT \texttt{o1} model was prompted to design a power factor correction scheme for the circuit in Fig.~\ref{fig:circuit_for_the_pf_correction_problem}. 

\begin{prompt}
The following problem for a university-level power systems engineering course explores power factor correction. Consider a balanced, single-phase, \(60\,\text{Hz}\) circuit with an AC voltage source with RMS value of \(V_{\mathrm{rms}} = 100\text{ V}\) connected to the parallel combination of a voltage source, a switched capacitor, and a load impedance with a complex power consumption of \(P_d + jQ_d\) VA. The amount of complex power supplied by the voltage source is denoted as \(P_s + jQ_s\) VA.
This switched capacitor bank consists of an inductor with reactance \(X_l\) in parallel with a capacitor with reactance \(X_c\) connected through switch \(S\). The inductor with reactance \(X_l\) is always connected in the circuit (not switchable).
The active power consumed by the load, \(P_d\), is a constant value of 50 W. The reactive power consumed by the load, \(Q_d\), continuously varies between a minimum value of -36 VAr and a maximum value of 60 VAr.
Your task in this problem is to determine the (fixed) values of the reactances \(X_l\) and \(X_c\) in Ohms as well as a switching strategy for the switch \(S\). Choose these so that the power factor of the power supplied by the voltage source (\(P_s+jQ_s\)) is as close to unity (1.0) as possible in terms of its magnitude (leading or lagging) for any value of the load’s reactive power demand \(Q_d\) within the range from -36 VAr (supplying 36 VAr) to 60 VAr (consuming  60 VAr).
\end{prompt}

\section{LLM Output, Errors, and Corrections}
\label{sec:solution}

When presented with the power factor correction prompt in Section~\ref{sec:prompt}, ChatGPT \texttt{o1} output an incorrect solution, which we describe below in Section~\ref{sec:chatgpt_sol}. Given the ChatGPT~\texttt{o1} output, students were tasked with identifying all errors, explaining the physical principle behind why it was incorrect, and ultimately providing a correct solution.

\subsection{ChatGPT's solution}
\label{sec:chatgpt_sol}
We next document the \emph{incorrect} output from ChatGPT~\texttt{o1} to the prompt in Section~\ref{sec:prompt}. Embedded in the lengthy solution proposed by ChatGPT, there are three locations where either critical engineering errors or fundamental misunderstandings of power systems were introduced. Each of these locations are demarcated with {\color{red}\raisebox{.5pt}{\textcircled{\raisebox{-.9pt} {1}}}}, {\color{red}\raisebox{.5pt}{\textcircled{\raisebox{-.9pt} {2}}}}, and {\color{red}\raisebox{.5pt}{\textcircled{\raisebox{-.9pt} {3}}}}, respectively.
\begin{response}
% \textit{Thought about power factor correction design for 25 seconds}

\noindent
\textbf{Analysis and Solution Strategy}

We have a single-phase, 60\,Hz source of $V_{\mathrm{rms}} \;=\; 100\,\text{V}$ supplying a load whose real power is fixed at $P_d \;=\; 50\,\text{W}$, and whose reactive power $Q_d$ can vary over the range $ -36\,\text{VAr} \;\le\; Q_d \;\le\; +60\,\text{VAr}.$

(Here, $Q_d<0$ means the load is supplying reactive power, i.e.\ effectively capacitive/leading, and $Q_d>0$ means the load is consuming reactive power, i.e.\ inductive/lagging.)

We wish to add a parallel branch of a \textbf{fixed} inductor (always connected) plus a \textbf{switchable} capacitor such that, no matter the value of $Q_d$ in that range, the net reactive power drawn from the $100$~V source is as close to zero as possible (i.e., power factor near unity). Denote:
\begin{itemize}
    \item The inductor's reactance by $X_l$.
    \item The capacitor's reactance by $X_c$.
    \item The switch $S$ either \textbf{opens} (removes the capacitor) or \textbf{closes} (places the capacitor in parallel).
\end{itemize}

Hence, depending on $S$, the net reactive power seen by the source is:
\begin{enumerate}
\item \textbf{Switch Open} (inductor only): $Q_s = Q_d + Q_L$ where $Q_L = \frac{V^2}{X_l} >0$ (inductive consumption).
\item \textbf{Switch Closed} (inductor + capacitor): $Q_s = Q_d + Q_L + Q_C$, where $Q_C = -\frac{V^2}{X_c}$ (capacitive supply).
\end{enumerate}

We want to choose $X_l$, $X_c$, and a switching ``rule'' so that $|Q_s| = |Q_d + Q_L + (\text{possibly~} Q_C)|$ is as small as possible for \textbf{all} $Q_d \in [-36,+60]$ VAr.

\vspace{0.5em}
\hrule
\vspace{0.5em}

\textbf{1. Match extremes at unity power factor.}

A common approach is to ensure that at the \textbf{two extreme} values of the load's reactive power, $Q_d = -36$~VAr (most leading) and $Q_d = +60$~VAr (most lagging), the source's net reactive power is zero in one of the two switch states. {\color{red}\raisebox{.5pt}{\textcircled{\raisebox{-.9pt} {1}}}}

\begin{enumerate}
\item \textbf{At \(Q_d = -36\,\text{VAr}\) (load is supplying 36 VAr):} \\
We would like to be able to correct $Q_s \approx 0$ \textbf{without} the capacitor, i.e. with the switch open.
$$0 = Q_s = Q_d + Q_L = -36 + \frac{100^2}{X_l}.$$

Solving for $X_l$:

$$\frac{10,000}{X_l} = 36 \Rightarrow X_l = \frac{10,000}{36} \approx 277.78~\Omega.$$

Thus the (always-connected) inductor's reactance is chosen to be $X_l \approx 278~\Omega$.

\item \textbf{At \(Q_d = +60\,\text{VAr}\) (load is consuming 60 VAr):} \\
We want to correct $Q_s \approx 0$ \textbf{with} the capacitor switched in. Then
$$0 = Q_s = Q_d + Q_L  + Q_C= 60 + \frac{10,000}{X_l} - \frac{10,000}{X_c}.$$

From the above we have $\frac{10,000}{X_l} = 36$. Hence
\begin{align*}
    60 + 36 - \frac{10,000}{X_c} = 0 &\Rightarrow \frac{10,000}{X_c} = 96\\
    &\Rightarrow X_c = \frac{10,000}{96} \approx 104.17~\Omega.
\end{align*}

Therefore, $X_l \approx 278~\Omega$, $X_c\approx 104~\Omega.$
\end{enumerate}

\vspace{0.5em}
\hrule
\vspace{0.5em}

\textbf{2. Reactive power vs. $Q_d$ under each switch state}

With the above reactances:
\begin{enumerate}
\item \textbf{Switch Open}: $Q_{s} = Q_d + 36$.
\begin{itemize}
    \item At $Q_d = -36$, indeed $Q_s = 0$.
    \item At $Q_d = 0$, $Q_s = +36$ VAr.
    \item At $Q_d = +60$, $Q_s = +96$ VAr.
\end{itemize}
\item \textbf{Switch Closed}: $Q_{s} = Q_d + 36 -96 = Q_d - 60$.
\begin{itemize}
    \item At $Q_d = -36$, indeed $Q_s = -96$ VAr.
    \item At $Q_d = 0$, $Q_s = -60$ VAr.
    \item At $Q_d = +60$, $Q_s = 0$.
\end{itemize}
\end{enumerate}

Graphically:
\begin{itemize}
    \item \textbf{Open}: $Q_s = Q_d + 36$, which is a straight line intercepting zero at $Q_d = -36$.
    \item \textbf{Closed}: $Q_s = Q_d -60$, intercepting zero at $Q_d = +60$.
\end{itemize}
% \vspace{0.2em}
\hrule
\vspace{0.2em}

\textbf{3. Best switching threshold}

We only have \textbf{two} discrete states (capacitor either in or out). At any intermediate $Q_d$, we cannot make $Q_s = 0$ exactly with a single fixed capacitor. Instead, we choose a \textbf{threshold} $Q_d^\ast$ for switching so as to keep $|Q_s|$ as small as possible overall.
\begin{itemize}
    \item When \textbf{Switch Open}, $Q_s = Q_d + 36$.
    \item When \textbf{Switch Closed}, $Q_s = Q_d - 60$.
\end{itemize}

We pick the state that yields the smaller magnitude $|Q_s|$. In other words:
\begin{enumerate}
    \item Compare $|Q_d + 36|$ vs. $|Q_d - 60|$.
    \item The "crossover" is where $|Q_d + 36| = |Q_d - 60|$.
    \begin{itemize}
        \item For $Q_d \geq 0$, we can drop absolute values and check
        $$Q_d + 36 = 60 - Q_d \Rightarrow 2 Q_d = 24 \Rightarrow Q_d = 12~\text{VAr}.$$
        
        \item Indeed for negative $Q_d$, the lines do not intersect in that region, but the best boundary for the entire range ends up around $Q_d = 12$ anyway.
    \end{itemize}
\end{enumerate}

Hence the switching "rule of thumb" is:
\begin{itemize}
    \item If $Q_d < 12$ VAr, use \textbf{Switch Open} (only inductor).
    \item If $Q_d > 12$ VAr, use \textbf{Switch Closed} (inductor + capacitor).
\end{itemize}

At $Q_d = 12$, either choice gives $|Q_s| = 48$ VAr. For values below 12, the open-switch magnitude is smaller; for values above 12, the closed-switch magnitude is smaller.

This simple 2-state approach places the net source reactive power $Q_s$ within at most $\pm 60$ or $\pm 96$ in the extremes {\color{red}\raisebox{.5pt}{\textcircled{\raisebox{-.9pt} {2}}}}, but it \textbf{exactly} hits zero at $Q_d = -36$ (switch open) and $Q_d = + 60$ (switched closed). One cannot do better with only a single fixed capacitor and a single fixed inductor. {\color{red}\raisebox{.5pt}{\textcircled{\raisebox{-.9pt} {3}}}}

\vspace{0.5em}
\hrule
\vspace{0.5em}

\textbf{4. Final numeric answers}

\begin{enumerate}
    \item \textbf{Inductor reactance}
    $$X_l = \frac{V^2}{36} = \frac{10,000}{36}\approx 277.8~\Omega $$

    At $f= 60$ Hz,
    $X_l = 2\pi f L \Rightarrow L = \frac{X_l}{2\pi f} \approx \frac{278}{2\pi \times 60} \approx 0.74~\text{H}.$

    \item \textbf{Capacitor reactance}
    $$ X_c = \frac{V^2}{96} = \frac{10,000}{96}\approx 104.2~\Omega$$

    At $f= 60$ Hz,
    $X_c = \frac{1}{2\pi f C} \Rightarrow C = \frac{1}{2\pi f X_c} \approx \frac{1}{2\pi \times 60 \times 104} \approx 25.5~\mu\text{F}.$
    
    \item \textbf{Switching strategy}
    \begin{itemize}
    \item \textbf{Switch Open} when $Q_d < 12$ VAr.
    \item \textbf{Switch Closed} when $Q_d > 12$ VAr.
    \item (Exactly at $Q_d=12$, either state yields $|Q_s| = 48$ VAr, so one might choose either.)
    \end{itemize}

    This ensures:
    \begin{itemize}
    \item Perfect cancellation $(Q_s = 0)$ at $Q_d = -36$ VAr (inductor only).
    \item Perfect cancellation $(Q_s = 0)$  at $Q_d = +60$ VAr (inductor + capacitor).
    \item For intermediate $Q_d$, the net reactive power from the source remains as small as possible given just these two discrete capacitor states.
     \end{itemize}

\end{enumerate}

% {\color{blue} \emph{The remainder of the output just repeats some of the information provided above as a ``summary".}}

\end{response}

\subsection{Analysis of ChatGPT's approach to power factor correction}
\label{sec:correcting_chatgpt}

ChatGPT's approach to this problem is somewhat reasonable but fundamentally incorrect. The idea of using the inductor's reactive power to ``shift'' the range of the reactive power supplied by the source and then correct the resulting power factor using a capacitor makes sense overall and is indeed the right intuition. Moreover, given ChatGPT's assertions about how to approach the problem, the numeric calculations it provides are mostly correct.

\subsubsection{Range of the reactive power compensation scheme}
\label{sec:q_ranges}
The output provided by ChatGPT breaks down at the point labeled by {\color{red}\raisebox{.5pt}{\textcircled{\raisebox{-.9pt} {1}}}} above. Here, ChatGPT decides to fix the supply's reactive power outputs to zero (unity power factor) at the extreme load demands, $Q_d = -36$ VAr and $Q_d = 60$ VAr. This is not the best possible approach to maximizing the smallest power factor, since it does not take advantage of the still large power factor that would occur immediately above and below the extreme values. In other words, with this strategy, the power factor would still be near unity for $Q_d = -37$ VAr and for $Q_d = 61$ VAr. This suggests that the solution could be improved by inward movement of the points where the supply provides zero reactive power so that these unity power factor points are within the range of $Q_d$ as opposed to being at the extremes.

\begin{figure}[t]
\centering
\begin{subfigure}{0.5\textwidth}
  \centering
  \begin{tikzpicture}
    \begin{axis}[width=\textwidth,
            height=2in,
            axis y line=middle,
            axis x line=middle,
            xlabel = $Q_d$,
            ylabel = $|Q_s|$,
            xtick={-36,12,60}, xmin=-40, xmax=64,
            ytick={0,48}, ymin=0, ymax=50,
            ylabel style={xshift=0pt,yshift=4pt},
            xlabel style={xshift=3pt},
            clip=false
            ]
        \addplot [domain=-36:12, 
            samples=100]{x+36};
        \addplot [domain=12:60, 
            samples=100]{-x+60};

        \node[yshift=0,xshift=35pt] at (axis cs:-23,-10) {\scriptsize $\leftarrow$ Switch Open};
        \node[yshift=0,xshift=-30pt] at (axis cs:45,-9.75) {\scriptsize Switch Closed $\rightarrow$};
    \end{axis}
  \end{tikzpicture}
  \caption{ChatGPT's Solution}
  \label{fig:sol_plot_chatgpt}
\end{subfigure}\\[0.3em]%
\begin{subfigure}{0.5\textwidth}
  \centering
    \begin{tikzpicture}
    \begin{axis}[width=\textwidth,
            height=2in,
            axis y line=middle,
            axis x line=middle,
            xlabel = $Q_d$,
            ylabel = $|Q_s|$,
            xtick={-36,-12,12,36,60}, xmin=-40, xmax=64,
            ytick={0,24,48}, ymin=0, ymax=50,
            clip=false
            ]
        \addplot [domain=-36:-12, 
            samples=100]{-x-12};
        \addplot [domain=-12:12, 
            samples=100]{x+12};
        \addplot [domain=12:36, 
            samples=100]{-x+36};
        \addplot [domain=36:60, 
            samples=100]{x-36};

        \node[yshift=0,xshift=-10pt] at (axis cs:0,-10) {\scriptsize $\leftarrow$ Switch Open };
        \node[yshift=0,xshift=10pt] at (axis cs:25,-9.75) {\scriptsize Switch Closed $\rightarrow$};
    \end{axis}
  \end{tikzpicture}
  \caption{Best Possible Solution}
  \label{fig:sol_plot_better}
\end{subfigure}
\caption{Visualization of ChatGPT's output and the best solution.}
\vspace{-1em}
\end{figure}
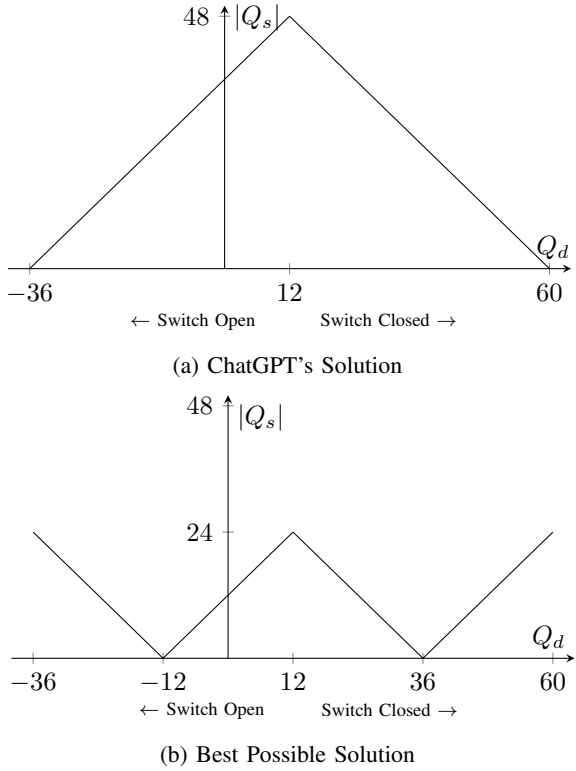

One can visualize this by considering the magnitude of the reactive power supplied or absorbed by the source, $|Q_s|$, as these unity power factor points change. When the capacitance and inductance values are chosen so that the unity power factor points are at the extremes of $Q_d = -36$ VAr and $Q_d = 60$ VAr, as in the ChatGPT output, $|Q_s|$ corresponds to the plot shown in Fig.~\ref{fig:sol_plot_chatgpt}, which has a maximum value of $|Q_s| = 48$~VAr. Conversely, moving these points towards the middle of the range 
% (say, to $Q_d = -30$ VAr and $Q_d = 42$ VAr) would give a plot like the one the right below, which has 
would yield a smaller maximum value of~$|Q_s|$.

The plot in Fig.~\ref{fig:sol_plot_better} has four line segments. The key insight is that the maximum smallest power factor will occur when each of these line segments has the same maximum height. Based on this intuition, the best possible solution evenly spaces the four line segments over the range of $Q_d$, as shown in Fig.~\ref{fig:sol_plot_better}.

% \begin{figure*}[!h]
% \centering
%   \centering
%   \begin{tikzpicture}
%     \begin{axis}[width=0.8\textwidth,
%             height=2in,
%             axis y line=middle,
%             axis x line=middle,
%             xlabel = $Q_d$,
%             ylabel = $|Q_s|$,
%             xtick={-36,-12,12,36,60}, xmin=-40, xmax=64,
%             % xticklabel={$\pgfmathprintnumber{\tick}^\circ$},
%             ytick={0,24,48}, ymin=0, ymax=50,
%             % yshift=\yshift,
%             clip=false
%             ]
%         \addplot [domain=-36:-12, 
%             samples=100]{-x-12};
%         \addplot [domain=-12:12, 
%             samples=100]{x+12};
%         \addplot [domain=12:36, 
%             samples=100]{-x+36};
%         \addplot [domain=36:60, 
%             samples=100]{x-36};

%         \node[yshift=0] at (axis cs:0,-10) {\scriptsize $\leftarrow$ Switch Open };
%         \node[yshift=0] at (axis cs:25,-9.75) {\scriptsize Switch Closed $\rightarrow$};
%     \end{axis}
%   \end{tikzpicture}
%   \caption{Best Possible Solution}
% \end{figure*}

In other words, the best possible solution places the points of unity power factor at one-fourth of the range of $Q_d$, that is, $Q_d = (60-(-36))\frac{1}{4} + (-36)= \frac{96}{4} - 36 = -12~\text{VAr}$ and three-fourths of the range of $Q_d$, that is, $Q_d = (60-(-36))\frac{3}{4} + (-36)= 72 - 36 = 36~\text{VAr}$, with the resulting maximum value of $|Q_s|$ being $24$~VAr which is reached at $Q_d = -36, 12$, and $60$ VAr. This curve results from switching the capacitor into the circuit at $Q_d = 12$ VAr.

\subsubsection{Corrected reactance values}
\label{sec:reactance}

We next need to choose appropriate values for $X_l$ and $X_c$. To select $X_l$, we observe that the power factor should be unity (and thus $Q_s$ should equal zero) when $Q_d = -12$ VAr. Since the switch is open at this point, only the inductor is in the circuit, so we have $Q_s = Q_d + \frac{V^2}{X_l}$. Plugging in the values for these quantities yields $0 = -12 + \frac{100^2}{X_l}$. Solving gives $X_l = 833.3~\Omega$. 
%As a sanity check, the inductor's reactance is positive, which is good since inductors must have positive reactances.

To compute $X_c$, we observe that the power factor should also be unity (and thus $Q_s$ should equal zero) when $Q_d = 36$ VAr. Since the switch is closed at this point, both the capacitor and inductor are in the circuit, and we thus have $Q_s = Q_d + \frac{V^2}{X_l} + \frac{V^2}{X_c}$. Substituting the values computed so far, we have $0 = 36 + \frac{100^2}{833.3} + \frac{100^2}{X_c}$. Solving gives $X_c = -208.3~\Omega$. 
%As another sanity check, $X_c$ is negative, which is what we would expect from a capacitor.

The smallest power factor occurs when $|Q_s|$ is at its maximum value of $24$~VAr, which happens when $Q_d = -36, 12$, and $60$ VAr. The value for this power factor at these points is given by $\frac{P_s}{\sqrt{P_s^2 + Q_s^2}}$. By conservation of power $P_s = P_d = 50$~W, so we have $\frac{50}{\sqrt{50^2 + 24^2}} = 0.9015$. The voltage source's power factor is leading when $Q_s > 0$ (source supplies reactive power), which occurs for values of $-12 < Q_d \leq 12$ and $36 < Q_d \leq 60$, and lagging when $Q_s < 0$ (source consumes reactive power), which occurs for $-36 \leq Q_d < -12$ and $12 < Q_d < 36$. This corrected solution is recorded in Tables~\ref{tab:corrections} and~\ref{tab:corrections_results}.
\begingroup
\renewcommand*{\arraystretch}{1.5}
\begin{table}[ht]
\centering
\caption{Corrected reactances}
\begin{tabular}{|>{\centering\arraybackslash} m{0.75cm}|| >{\centering\arraybackslash} m{2.5cm}|||>{\centering\arraybackslash} m{0.75cm}|| >{\centering\arraybackslash} m{2.5cm}|}
\hline  $X_l$& { $833.3~\Omega$} &   $X_c$& { $-208.3~\Omega$} \\
\hline
\end{tabular}
\label{tab:corrections}
\end{table}
\endgroup
\vspace{-1.25em}
\begingroup
\renewcommand*{\arraystretch}{1.5}
\begin{table}[ht]
\centering
\caption{Corrected reactive power compensation}
\resizebox{\linewidth}{!}{%
\begin{tabular}{|>{\centering\arraybackslash} m{0.75cm}|| >{\centering\arraybackslash} m{2.5cm}| >{\centering\arraybackslash} m{4.5cm}|}
\hline   \textbf{Switch Status} &   \textbf{Range of $\mathbf{Q_d}$} &   \textbf{Smallest Source Power Factor}\\
\hline   Open &   $-36 \leq Q_d \leq 12$ VAr  &   0.9015 (lagging when $Q_d = -36$ VAr, leading when $Q_d = 12$ VAr)\\
\hline   Closed &   $12 < Q_d \leq 60$ VAr  &   0.9015 (lagging when $Q_d = 12$ VAr,\newline leading when $Q_d = 60$ VAr)\\
\hline
\end{tabular}}
\label{tab:corrections_results}
\end{table}
\endgroup

ChatGPT makes two other errors in its solution. At the point labeled by {\color{red}\raisebox{.5pt}{\textcircled{\raisebox{-.9pt} {2}}}} above, the numeric value is incorrect. With the $X_l$ and $X_c$ values and the switching strategy proposed by ChatGPT, the values of $Q_s$ will range between $0$ and $48$, not $\pm 60$ or $\pm 96$ as indicated in the ChatGPT solution.

Finally, the statement that ``one cannot do better with only a single fixed capacitor and a single fixed inductor'' at the point labeled by {\color{red}\raisebox{.5pt}{\textcircled{\raisebox{-.9pt} {3}}}} is meaningless as we have a \emph{switched} capacitor.
%, not a fixed capacitor.

\section{Data and Discussion Regarding Student Success}
\label{sec:results}
As a retrospective analysis, we now describe the success of ECE 4320 students at analyzing errors in the ChatGPT output.

\subsection{Spring 2025: Student success in identifying errors in the output of ChatGPT o1}
\label{sec:student_success}
Students had varying levels of success in identifying ChatGPT's errors in the power factor correction problem from Section~\ref{sec:prob-form}. The grading rubric results summarized in Table~\ref{tab:student_success} show the number of students who identified three, two, one, or zero of ChatGPT's errors.

While a plurality of students identified all three ChatGPT errors, a substantial majority ($66.33\%$) did not detect all errors. The outcome that was the most concerning, in our view, was that approximately a quarter of the class did not detect \emph{any} of the errors. Anecdotally, one student mentioned first solving the problem themselves without looking at the ChatGPT output, getting the same answer as that provided by ChatGPT (known to be incorrect), and then being unsure how to proceed. 

\begin{table}[t]
\centering
\caption{Distribution of class success in identifying power engineering mistakes made by ChatGPT \texttt{o1} ($n=30$).}
\begin{tabular}{l||c|c}
\textbf{\# of Errors Identified} & \textbf{Pct. of Students} & \textbf{\# of Students} \\ \hline
3/3                                         & 36.66\%    &    11       \\
2/3                                         & 16.66\%    &    5        \\
1/3                                         & 23.33\%    &    7         \\
0/3                                         & 23.33\%    &    7  \\
\hline
\end{tabular}
\label{tab:student_success}

\end{table}

\subsection{Student success in correcting ChatGPT's errors}
\label{sec:corrections}

The second part of the question asked students to fix the ChatGPT errors by providing the correct solution. As summarized by the grading rubric results in Table \ref{tab:student_correction}, these outcomes also varied. A marginal majority of the class ($\approx 51 \%$) successfully corrected the key errors, up to minor issues in the range of $Q_d$ or the reactances. Conversely, about a quarter of the class failed to fully correct any of the errors. While roughly consistent with the number of students that did not identify any of the ChatGPT errors in ChatGPT, note that these were not all the same students.

\begin{table}[t]
\centering
\caption{Distribution of class success in correcting the power engineering mistakes made by ChatGPT \texttt{o1} $(n=30)$.}
\begin{tabular}{l||c|c}
\textbf{Concepts corrected}        &  \textbf{Pct. of Students} & \textbf{\# of Students} \\\hline
$Q_d$, $X_l$, $X_c$, and power factor & 27\%                  & 8  \\
$X_l$, $X_c$, and power factors       & 24\%                  & 7   \\
$Q_d$ only              & 10\%      & 3\\
Partially, more than half                & 20\%            & 6   \\
Partially, less than half               & 14\%            & 4 \\
None                          & \hphantom{0}7\% & 2\\
\hline
\end{tabular}
\label{tab:student_correction}
\vspace{-1em}
\end{table}

% \subsection{Crowdsourcing Problems that Challenge ChatGPT}

\subsection{Contrasting with Fall 2023: GPT-4}
\label{sec:gpt4}
In the Fall 2023 semester, we used a predecessor to \texttt{o1}, ChatGPT version \texttt{GPT-4}, for a similar power factor correction problem with variations in the circuit parameters. As discussed in the appendix, the ChatGPT \texttt{GPT-4} output in Fall 2023 included nearly a dozen significant errors, many of which were much more obvious than the relatively subtle logical error from ChatGPT \texttt{o1} in Spring 2025. Accordingly, students in the Fall 2023 semester performed significantly better at error identification, with 21 of 28 students (75\%) finding all or nearly all errors. However, students still found error correction to be challenging, with 18 of 28 students (64\%) failing to provide a fully correct solution.

% Fast moving field

\subsection{Discussion and limitations}
\label{sec:discussion}

These results suggest that logical errors made by recent versions of LLMs may often be conceptually similar to the errors made by students. Thus, educators may find it difficult to determine whether a student's erroneous solution is simply copied from an LLM or instead indicates a student's misunderstanding in independent work. Moreover, given the rapid quality improvement between ChatGPT version \texttt{GPT-4} in Fall 2023 and \texttt{o1} in Spring 2025 (as well as subsequent improvements in \texttt{o3}; see Appendix~\ref{apdx:new_models}), educators should keep up to date with evolving LLM capabilities and typical errors.

Understanding the underlying reasons for the similarity between student mistakes and LLM output errors could give insights into LLM characteristics with respect to power engineering. Since we independently created the power factor problems presented in this paper, these problems could not be part of any LLM training datasets and thus the LLMs are not simply regurgitating training data containing erroneous student solutions. This suggests a deeper justification for this similarity.

We also note several limitations with findings in this paper. The results presented in Section~\ref{sec:results} are retrospective in nature; we only report our experiences in selected assignments for this specific course. The results shown in these tables do not indicate or imply a distribution of the ability of the more general power engineering student population to identify or correct LLM errors. We also do not further describe the precise methodologies used by the students to protect student privacy. This motivates further studies on LLMs that consider additional power engineering courses and different types of problems.

\section{Conclusion}
\label{sec:conclusion}

As LLMs advance, their role in higher education requires careful examination. This is particularly the case in fields that impact real-world systems, such as electric power engineering. This study assessed the performance of ChatGPT in an undergraduate power system engineering course. The results show that many students struggled to identify errors in the ChatGPT outputs. The share of the class that was unable to identify any of ChatGPT's errors was larger than the share that was able to solve the entire problem correctly themselves. These findings highlight the need for robust assessment strategies and AI-aware grading policies in power engineering education.

{\footnotesize
\bibliographystyle{ieeetr}
\bibliography{refs}
}
\appendix[Contrasting with Fall 2023: The Pitfalls of GPT-4]

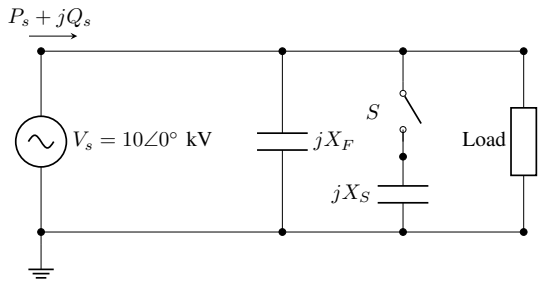
\begin{figure}[t]
\centering
\scalebox{0.802}{%
\begin{circuitikz}[american currents,european resistors]
  %% Source
  \coordinate (Csource);
  \draw +({90}:3) coordinate (Sa) node[anchor={0}]{} to[vsourcesin,l^=\mbox{$V_s = 10\angle 0^\circ$~kV},*-*] (Csource);
  
  \node (Sg) at ($(Csource)+(0.15,3.55)$){$P_s + j Q_s$};
  \path [->,>=stealth] ($(Csource)+(-0.2,3.25)$) edge ($(Csource)+(0.6,3.25)$);
  
  %% Loads
  \coordinate (Cload1) at ($(Csource)+(8,0)$);
  \draw (Cload1) to[R,l^=\mbox{Load},*-*] +({0+90}:3) coordinate (La) node[anchor=south]{};

  %% Fixed capacitor
  \coordinate (FC1) at ($(Csource)+(4,0)$);
  \draw (FC1) to[C,l_=$jX_{F}$,*-*] +({0+90}:3) node[anchor=south]{};

  %% Switched Capacitor
  \switchedCapacitor{SC1}{$(Csource)+(6,0)$}{$jX_{S}$}{$S$}
 
  % %% Connections
  \draw (Sa) -- +(0,0) to[short] ($(La)+(0,0)$) -- (La);
  \draw (0,0) -- ($(La)+(0,-3)$);
  
  \draw ($(Csource)+(0,0)$) -- +(0,-0.2) node[ground]{};

\end{circuitikz}}
\caption{The circuit presented to the Fall 2023 cohort of ECE~4320 students and ChatGPT \texttt{GPT-4}.}
\label{fig:gpt4-circuit}
% \vspace{-1em}
\end{figure}

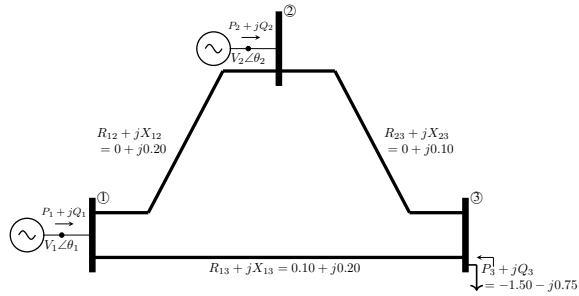
\begin{figure}[h!]
% \centering
\scalebox{0.62}{%
\begin{circuitikz}[american currents,european resistors, scale = 0.8, transform shape]
\ctikzset{bipoles/length=1.5cm}

% \bus: Bus component for a one-line diagram.
% Args:
% #1: Name of component
% #2: Center coordinates
% #3: Bus number
% #4: Bus number location (UL,UR,LL,LR for upper left, upper right, lower left, lower right)
% #5: Load demand left (set to 0 to draw the arrow without a label)
% #6: Load demand right
% Note: Multiple terminals are available (evenly spaced, listed from top to bottom):
%.        Two terminals: (#1_2.1), (#1_2.2)
%.        Three terminals: (#1_3.1), (#1_3.2), (#1_3.3)
%.        Four terminals: (#1_4.1), (#1_4.2), (#1_4.3), (#1_4.4)
%.        Four terminals: (#1_5.1), (#1_5.2), (#1_5.3), (#1_5.4), (#1_5.5)
\bus{B1}{(0,1)}{1}{UR}{}{}
\bus{B2}{($(B1)+(5,5)$)}{2}{UR}{}{}
\bus{B3}{($(B1)+(10,0)$)}{3}{UR}{}{$P_3 + jQ_3$\\ \;$= -1.50 - j0.75$}

\draw ($(B3)+(0.75,-0.85)$) -- ++(0,0.25);
\path [->,>=stealth] ($(B3)+(0.75,-0.85)+(0,0.25)$) edge ++(-0.45,0);

\phantom{\bus{B1}{(B1)}{3}{UR}{$P_3 + jQ_3 = -1.50 - j0.75$}{}} % Add a phantom load on the left to balance out the spacing

% \Sline: S shaped line between specified terminal locations.
% Args:
% #1: Name of component
% #2: First terminal location
% #3: Second terinal location
% #4: Horizontal offset
% #5: Vertical offset
% #6: Impedance label
% #7: Impedance label location shift
% #8: First breaker label (leave empty to not have a breaker or 0 to have a breaker without a label)
% #9: Second breaker label (leave empty to not have a breaker or 0 to have a breaker without a label)
% Note: This function assumes the first bus is to the left of the second bus.
\Sline{L12}{B1_3_1}{B2_3_3}{1.5}{0}{$R_{12}+jX_{12}$\\ \;$=0+j0.20$}{(-2.5,0)}{}{}
\Sline{L23}{B2_3_3}{B3_3_1}{1.5}{0}{$R_{23}+jX_{23}$\\ \;$=0+j0.10$}{(0.2,0)}{}{}

% \straightline: Straight line between specified terminal locations
% Args:
% #1: Name of component
% #2: First terminal location
% #3: Second terinal location
% #4: Impedance label
% #5: Impedance label location shift
% #6: First breaker label (leave empty to not have a breaker or 0 to have a breaker without a label)
% #7: Second breaker label (leave empty to not have a breaker or 0 to have a breaker without a label)
\straightline{L13}{B1_3_3}{B3_3_3}{$R_{13} + jX_{13} = 0.10 + j0.20$}{(-2,-0.3)}{}{}

% \generatorL: Generator component for a one-line diagram, oriented left to right.
% Args:
% #1: Name of component
% #2: Center coordinates
% #3: Grounding connection (either "delta" or "wye" for an ungrounded wye or a number indicating the grounding impedance for a wye connection, with zero interpreted as a short to ground)
% #4: Top label
% #5: Internal reactance label
% #6: Power injection arrow label
% #7: Terminal voltage label
% #8: Hide node dot at connection point
% Note: Terminal is available as (#1_right).
\generatorL{G1}{($(B1_3_2)+(-1.75,0)$)}{}{}{}{$P_1+jQ_1$}{V_1\angle \theta_1}{1}
\draw (G1_right) -- (B1_3_2);

\generatorL{G2}{($(B2_3_2)+(-1.75,0)$)}{}{}{}{$P_2+jQ_2$}{V_2\angle \theta_2}{1}
\draw (G2_right) -- (B2_3_2);

\end{circuitikz}}
\caption{The one-line diagram presented to the Fall 2023 cohort of ECE 4320 students and ChatGPT \texttt{GPT-4}.}
\label{fig:gpt4-oneline}
\end{figure}

In the Fall 2023 semester, we experimented with the \texttt{GPT-4} model on a similar power factor correction problem as well as a power flow formulation problem, using the circuits shown in Fig. \ref{fig:gpt4-circuit} and Fig. \ref{fig:gpt4-oneline}. The problem settings presented to students and \texttt{GPT-4} are shown in the prompts below, with corrections.

\subsection{Power factor correction problem}
\label{apdx:gpt-4-pf-correction}

% This power factor correction problem focuses on the single-phase, 60~Hz circuit shown below in Fig. \ref{fig:gpt4-circuit}.

\begin{prompt}
    The following problem for a university-level power systems engineering course explores power factor correction. Consider a balanced, single-phase, 60 Hz circuit with a 10~kV AC voltage source connected to the parallel combination of a load, a capacitor, and a switched capacitor. The load consumes different amounts of power over time. In the morning, the load consumes complex power of 10 + j5 MVA. In the afternoon, the load consumes apparent power of 40 MVA at a power factor of 0.8 lagging. In the evening, the load consumes real power of 10 MW at a power factor of unity (1.0). The fixed capacitor is always connected in parallel with the load and the voltage sources, while the switched capacitor can be switched into or out of the circuit. The capacitors' reactances do not change across time periods.

    Consider the power factor of the power supplied by the voltage source. Your task for this problem is to choose 1) the best values of capacitive reactances for both the fixed capacitor and the switched capacitor and 2) the switching configurations for the switched capacitor during each time period (morning, afternoon, and evening) that maximize the smallest power factor across the day. In other words, choose the capacitors' reactances and the switching configurations such that the smallest power factor during any time period is as large as possible (close to 1, either leading or lagging). What capacitive reactances would you select, what switching configurations would you choose during each time period, and what are the corresponding power factors? Indicate whether these power factors are leading or lagging.
\end{prompt}
\vspace{-1em}
\begin{response}
    To maximize the smallest power factor across the day, we need to determine the 
capacitive reactances and switching configurations for each time period. Let's 
analyze each time period separately:
\textbf{Morning:} The load consumes complex power of $10 + j5$ MVA. To maximize the smallest power factor, 
we want the power factor to be close to 1 (leading or lagging). \textcolor{red}{Since the load's power factor is not specified~\raisebox{.5pt}{\textcircled{\raisebox{-.9pt} {1}}}}, we can choose a leading power factor for simplicity. Let's aim for a power factor of 0.9 leading.

We can calculate the apparent power ($S$) using the complex power ($S = P + jQ$). Given
that the complex power is $10 + j5$ MVA, \textcolor{red}{the apparent power is 10 MVA.~\raisebox{.5pt}{\textcircled{\raisebox{-.9pt} {2}}}} Since the 
voltage is 10 kV, the \textcolor{red}{load impedance is $Z = V^2 / S = (10^4)^2 / 10^7 = 10~\Omega$.~\raisebox{.5pt}{\textcircled{\raisebox{-.9pt} {3}}}}

Let's select the fixed capacitor's reactance ($X_{c,\text{fixed}}$) such that it compensates the load's reactive power ($Q$) and provides the desired leading power factor. \textcolor{red}{The reactance can be calculated as $X_{c,\text{fixed}} = -Q / (2\pi f)$, where $f$ is the frequency.}
$${\color{red} X_{c,\text{fixed}} = -5~\text{MVAR} / (2\pi * 60~\text{Hz}) = -13.2~\Omega~\raisebox{.5pt}{\textcircled{\raisebox{-.9pt} {4}}}}$$
\textcolor{red}{Since the reactance should be capacitive, we can use $X_{c,\text{fixed}} = 13.2~\Omega$.~\raisebox{.5pt}{\textcircled{\raisebox{-.9pt} {5}}}}

Since the switched capacitor does not need to be used during the morning period (since 
the fixed capacitor already provides the desired power factor), we do not need to 
consider the switching configuration.

\textbf{Afternoon:}
The load consumes apparent power of 40 MVA at a power factor of 0.8 lagging. To
maximize the smallest power factor, we need to bring this power factor closer to
1 (leading or lagging). 
\end{response}

\begin{response}
\textcolor{red}{Let's aim for a power factor of 0.9 lagging.~\raisebox{.5pt}{\textcircled{\raisebox{-.9pt} {6}}}}
The load's real power ($P$) can be calculated as $P = S * \text{power factor} = 40~\text{MVA} * 0.8 = 32~\text{MW}$. The reactive power ($Q$) can be determined using $Q = \sqrt{S^2 - P^2} = \sqrt{(40~\text{MVA})^2 - (32~\text{MW})^2} = 24~\text{MVAR}$.

Using the \textcolor{red}{load impedance formula $Z = V^2 / S$, we can find the load impedance as $Z = (10^4)^2 / 40~\text{MVA} = 2500~\Omega$.~\raisebox{.5pt}{\textcircled{\raisebox{-.9pt} {7}}}}

To compensate for the reactive power and adjust the power factor to 0.9 lagging, we need to find the required capacitive reactance. The reactive power can be calculated as $Q = V^2 / X_c$, where $X_c$ is the capacitive reactance.
$${\color{red}X_c = V^2 / Q = (10^4)^2 / 24~\text{MVAR} = 4167~\Omega~\raisebox{.5pt}{\textcircled{\raisebox{-.9pt} {8}}}}$$
Since we already have the fixed capacitor with reactance $X_{c,\text{fixed}} = 13.2~\Omega$, we need to find the switched capacitor's reactance $X_{c,\text{switched}}$) such that the \textcolor{red}{total capacitive reactance is equal to the required $X_c - X_{c,\text{switched}}$.}
$${\color{red}X_{c,\text{switched}} = X_c - X_{c,\text{fixed}} = 4167~\Omega - 13.2~\Omega = 4154~\Omega~~\raisebox{.5pt}{\textcircled{\raisebox{-.9pt} {9}}}}$$
The switching configuration for the switched capacitor during the afternoon period is to connect it in parallel with the fixed capacitor and the load.

\textbf{Evening:}
The load consumes real power of 10 MW at a power factor of unity (1.0). \textcolor{red}{Since the power factor is already 1.0, there is no need for power factor correction during this period. Therefore, we don't need to consider capacitors or switching configurations for the evening period.~\raisebox{.5pt}{\textcircled{\raisebox{-.7pt} {\footnotesize 10}}}}

\textbf{Summary:}
To maximize the smallest power factor across the day, the chosen values for the capacitive reactances and switching configurations are as follows:

{\color{red}\textbf{Morning:}
Fixed capacitor: $X_{c,\text{fixed}} = 13.2~\Omega$ (connected in parallel with the load and voltage source)
Switched capacitor: Not used

\textbf{Afternoon:}
Fixed capacitor: $X_{c,\text{fixed}} = 13.2~\Omega$ (connected in parallel with the load and voltage source)
Switched capacitor: $X_{c,\text{fixed}} = 4154~\Omega$ (connected in parallel with the load, fixed capacitor, and voltage source)

\textbf{Evening:}
No power factor correction is required.

The corresponding power factors for each time period are:
Morning: 0.9 leading power factor
Afternoon: 0.9 lagging power factor
Evening: 1.0 (unity power factor)~\raisebox{.5pt}{\textcircled{\raisebox{-.7pt} {\footnotesize 11}}}}
\end{response}

In contrast with \texttt{o1}, the \texttt{GPT-4} response exhibited errors of a much more fundamental character, and in significantly greater number. Below, we summarize these errors in the order in which they are marked in the above response.

\vspace{0.5em}

{\scriptsize
\begin{enumerate}[label=\arabic*)]
    \item The load's power factor is not given during the morning time period, but we can infer this from the complex power consumption of $10+j5$~MVA. Specifically, the power factor is $10/\sqrt{10^5 + 5^2} = 0.8944$ lagging. There is no need to arbitrarily choose a power factor for the load during this or any other time period.

    \item With a complex power of $10 + j5$~MVA, the apparent power is $\sqrt{10^2 + 5^2} = 11.1803$~MVA.

    \item The formula $Z = V^2 / S$ is not correct. The correct expression is $Z = |V|^2 / S^\ast$. The load impedance is $Z =(10\times 10^3)^2 / (10\times 10^6 + j5\times 10^6)^\ast = 8+j4~\Omega$.

    \item The reactive power consumed by a shunt capacitance is $Q = -|V|^2/(X_c)$, so the capacitive reactance that would supply a 5~MVAr is $X_{c,\text{fixed}} = -|V|^2 / Q = -(10\times 10^3)^2 / 5\times 10^6 = -20~\Omega$, not $-13.2~\Omega$. Additionally, we do not necessarily want the fixed capacitor to supply 5~MVAr. While this will lead to unity power factor in the morning, this much reactive power will lead to an oversupply of reactive power during the evening.

    \item A capacitor has a negative reactance, so $X_{c,\text{fixed}}$ should be a negative value.

    \item The assertion that we should aim for a power factor of 0.9 lagging is not correct. We can choose values that result in a higher power factor during all time periods.

    \item Same as in 4), the formula $Z = V^2 / S$ is not correct (and the value is also wrong even if the formula had been correct). The load impedance during the afternoon should be $Z =(10\times 10^3)^2 / (32\times 10^6 + j24\times 10^6)^\ast = 2+j1.5~\Omega$.

    \item This expression for the desired capacitive reactance has both a sign error and the value of $4167~\Omega$ is not the output of this expression (off by three orders of magnitude). The value should be $(10\times 10^3)^2 / -24\times 10^6 = -4.1667~\Omega$.

    \item The switched capacitor's reactance is based on the difference between the fixed capacitor's reactance and the desired reactance. However, it is not correct to directly subtract the reactances since the capacitors are connected in parallel. %Instead, we should compute the desired reactive power from the switched capacitor as the difference between the total reactive power compensation and the reactive power supplied by the fixed capacitor and then use the resulting value to find the correct reactance for the switched capacitor.

    \item This analysis of the evening load demand neglects the fact that the fixed capacitor is supplying reactive power and thus changing the power factor away from unity. The choice of the fixed capacitor's reactance should be cognizant of the resulting deviation in the power factor during the evening period.

    \item The values in the summary are incorrect due to the errors earlier in the solution.
\end{enumerate}}

%%%%%%%%%%%%%%%%%%%%%%%%%%
\vspace{-1em}
\subsection{AC power flow formulation problem}
\label{apdx:gpt4-ac-pf}

% Consider the three-bus system with the one-line diagram below. The lines are modeled as series impedances with the values labeled on the diagram in per unit.

% Bus~3 has a load that consumes a specified complex power of $1.50 + j0.75$~pu, corresponding to a complex power injection into the network of $P_3 + jQ_3 = -1.5 - j0.75$~pu.

% To implement a so-called ``distributed slack'' formulation, buses~1 and~2 are different types of buses than we have encountered so far. Bus~1 sets the reference angle by specifying $\theta_1 = 0^\circ$. The generators at both buses~1 and~2 maintain constant voltage magnitudes at their respective buses of 1.0 per unit; in other words, $V_1 = 1.0$ per unit and $V_2 = 1.0$ per unit. Furthermore, the generators at buses~1 and~2 equally share responsibility for supplying the load demand at bus~3 and the losses in the transmission lines by specifying that $P_1 = P_2$.

% This problem involves writing the power flow equations for this system by listing the variables in a vector~$x$ and the vector of equations $f(x)$ in the form $f(x) = 0$. Your expressions in $f(x)$ should be written solely in terms of numeric values and the variables in your vector~$x$. In other words, you should substitute in all parameters with numeric values. 

\begin{prompt}
    The following problem for a university-level power systems engineering course explores the formulation of a power flow problem with non-standard bus types. Consider a balanced, single-phase, 60 Hz power system with three buses. Bus 1 is connected to bus 2 via a transmission line with a series impedance of $R_{12} + jX_{12} = 0 + j0.20$. Bus 1 is also connected to bus 3 via a transmission line with a series impedance of $R_{13} + jX_{13} = 0.10 + j0.20$. Bus 2 is connected to bus 3 via a transmission line with a series impedance of $R_{23} + jX_{23} = 0 + j0.10$. 

    Bus 3 is a PQ bus modeling a load that consumes real power $P_3 = 1.50$ per unit and reactive power of 0.75 per unit. Bus 1 and bus 2 have generators that both maintain voltage magnitudes of 1.0 per unit at their respective buses, i.e., $V_1 = 1.0$ per unit and $V_2 = 1.0$ per unit. Bus 1 sets the voltage angle reference such that $\theta_1 = 0$ degrees. The generators at bus 1 and bus 2 equally share responsibility for balancing the active power in the system via a distributed slack bus formulation such that $P_1 = P_2$.
    
    Your task for this problem is to write the power flow equations for this system. List all variables for this problem in a vector x and all equations in a vector $f(x)$ such that $f(x) = 0$. Provide numeric values for all coefficients in the equations. Express all equations with real-valued quantities.
\end{prompt}

% When prompted with the text in the box above, \texttt{GPT-4} provided the following \textbf{incorrect} solution.
\vspace{-1em}
\newcounter{err}
\begin{response}
    %%%% Here is the ChatGPT4 response
To formulate the power flow equations for the given power system, we need to define the variables and equations based on the information provided. Let's use the following notation:

Variables (per unit):
% \vspace{-1em}
\begin{itemize}[itemsep=0em]
    \item {\color{red}$V_1$: Voltage magnitude at bus 1}
    \item {\color{red}$V_2$: Voltage magnitude at bus 2~\stepcounter{err}\raisebox{.5pt}{\textcircled{\raisebox{-.9pt} {\theerr}}}}
    \item $V_3$: Voltage magnitude at bus 3
    \item $\theta_2$: Voltage angle at bus 2 (with respect to bus 1)
    \item $\theta_3$: Voltage angle at bus 3 (with respect to bus 1)
\end{itemize}

Variables (real and reactive powers in per unit):
% \vspace{-1em}
\begin{itemize}[itemsep=0em]
    \item $P_1$: Real power generated at bus 1
    \item $Q_1$: Reactive power generated at bus 1
    \item $P_2$: Real power generated at bus 2
    \item $Q_2$: Reactive power generated at bus 2
    \item {\color{red}$P_3$: Real power consumed at bus 3}
    \item {\color{red}$Q_3$: Reactive power consumed at bus 3~\stepcounter{err}\raisebox{.5pt}{\textcircled{\raisebox{-.9pt} {\theerr}}}}
\end{itemize}

The system has three buses, and each bus has its own power balance equation:

Bus 1 (Slack bus with distributed slack formulation for active power balance):
% \vspace{-1em}
\begin{itemize}[itemsep=0em]
    \item $P_1 = P_2$
\end{itemize}

Bus 2 {\color{red}(PQ bus with specified real and reactive power injections):~\stepcounter{err}\raisebox{.5pt}{\textcircled{\raisebox{-.9pt} {\theerr}}}}
% \vspace{-1em}
\begin{itemize}[itemsep=0em]
    \item {\color{red}$P_2 - V_2 \cdot V_1\cdot (0.20\cdot \sin(\theta_2) + 0.10 \cdot \sin(\theta_2 - \theta_3)) = 0$}
    \item {\color{red}$Q_2 - V_2 \cdot V_1 \cdot (0.20 \cdot \cos(\theta_2) + 0.10 \cdot \cos(\theta_2 - \theta_3)) = 0$~\stepcounter{err}\raisebox{.5pt}{\textcircled{\raisebox{-.9pt} {\theerr}}}}
\end{itemize}

\pagebreak[4]
Bus 3 (PQ bus with specified real and reactive power consumptions):
% \vspace{-1em}
\begin{itemize}[itemsep=0em]
    \item {\color{red}$P_3 - V_3 \cdot V_1 \cdot 0.10 \cdot \sin(\theta_3) = 0$}
    \item {\color{red}$Q_3 - V_3 \cdot V_1 \cdot 0.10 \cdot \cos(\theta_3) = 0$~\stepcounter{err}\raisebox{.5pt}{\textcircled{\raisebox{-.9pt} {\theerr}}}}
\end{itemize}

\end{response}

\begin{response}
Additionally, the voltage magnitudes are fixed at buses 1 and 2:

Voltage magnitude constraints:
% \vspace{-1em}
\begin{itemize}[itemsep=0em]
    \item $V_1 - 1.0 = 0$
    \item $V_2 - 1.0 = 0$
\end{itemize}
With these equations, we can now write the system of equations $f(x) = 0$:
% \vspace{-1em}
\begin{align*}\color{red}
f(x) = \begin{bmatrix}    
& P_1 - P_2\\
& P_2 - V_2 \cdot V_1 \cdot (0.20 \cdot \sin(\theta_2) + 0.10 \cdot \sin(\theta_2 - \theta_3))\\
& Q_2 - V_2 \cdot V_1 \cdot (0.20 \cdot \cos(\theta_2) + 0.10 \cdot \cos(\theta_2 - \theta_3))\\
& P_3 - V_3 \cdot V_1 \cdot 0.10 \cdot \sin(\theta_3)\\
& Q_3 - V_3 \cdot V_1 \cdot 0.10 \cdot \cos(\theta_3)\\
& V_1 - 1.0\\
& V_2 - 1.0
\end{bmatrix}~\stepcounter{err}\raisebox{.5pt}{\textcircled{\raisebox{-.9pt} {\theerr}}}
\end{align*}

The vector $x$ contains all the variables:
\begin{equation*}\color{red}
x = \begin{bmatrix}V_1\\ V_2\\ V_3\\ \theta_2\\ \theta_3\\ P_1\\ Q_1\\ P_2\\ Q_2\\ P_3\\ Q_3\end{bmatrix}~\stepcounter{err}\raisebox{.5pt}{\textcircled{\raisebox{-.9pt} {\theerr}}}
\end{equation*}

The coefficients for the power flow equations are:
% \vspace{-1em}
\begin{itemize}[itemsep=0em]
    \item The coefficients for voltage magnitudes $(V_1, V_2, V_3)$ are all 1.0.
    \item The coefficients for voltage angles ($\theta_2$, $\theta_3$) are not explicitly provided but are in the trigonometric functions.
    \item The coefficients for real and reactive powers are either 0 or 1, and {\color{red}some are multiplied by impedance values, which are given in the problem description.~\stepcounter{err}\raisebox{.5pt}{\textcircled{\raisebox{-.9pt} {\theerr}}}}
\end{itemize}

Please note that this is a simplified per unit power flow model, and {\color{red}other practical factors like line losses, tap ratios, and shunt elements are not included.~\stepcounter{err}\raisebox{.5pt}{\textcircled{\raisebox{-.9pt} {\theerr}}}}
\end{response}

The errors in the \texttt{GPT-4} output are indicated with red highlighting in the text box above. Below, we discuss these errors in the order in which they appear in the response.

\vspace{0.5em}

{\scriptsize
\begin{enumerate}[label=\arabic*)]
    \item The voltage magnitudes at buses~1 and~2 are specified values, so these are not variables. The ChatGPT solution adds constraints to enforce $V_1 = 1$ and $V_2 = 1$, which is not wrong but is not necessary since these can just be eliminated from the problem by substituting in the corresponding values.
    
    \item The active and reactive power injections at bus~3 are specified values, so these are not variables.

    \item Bus~2 is not a PQ bus since neither the active power nor the reactive power are specified. Rather, the voltage magnitude $V_2$ is specified and the active power, while not specified explicitly, is equal to the active power at bus~1. 
    
    \item The expressions for active and reactive power balance at bus~2 have several errors. First, these expressions are missing the terms corresponding to the power flowing on the line from bus~2 to bus~3. Second, the expressions use the line resistance and reactance values instead of the conductance and susceptance values. Third, the reactive power balance equation is not needed since Q2 is not a specified quantity (can take any value such that the other equations are satisfied).

    \item The expressions for active and reactive power balance at bus~3 have several errors. First, the expressions for active and reactive power balance at bus 3 are missing the specified values for $P_3 = -1.50$ and $Q_3 = -0.75$. Second, these expressions are missing the terms corresponding to the line between buses~2 and~3. Third, the expressions use the line resistance value instead of the conductance and susceptance values.

    \item The specified $x$ vector and system of equations for $f(x) = 0$ inherit the errors above and also are clearly incorrect since the number of equations (seven) is inconsistent with the number of variables (eleven).

    \item See \textcircled{6} above.

    \item The real and reactive power expressions are not multiplied by the impedance values. Actually, this whole portion of the response is not needed.

    \item The problem does not have any transformers, so tap ratios do not need to be considered, and there are also no shunt elements. Finally, the formulation (if appropriately written) should account for line losses.
\end{enumerate}
}

\newpage
\subsection{Can new LLM models do better?}
\label{apdx:new_models}
At the conclusion of the Spring 2025 semester, the \texttt{o3} reasoning model was released by OpenAI. This section summarizes the performance of these new advanced reasoning models on all the preceding power engineering problems. In stark contrast to previous models, \texttt{o3} was essentially correct for both power factor correction problems; however, it continued to struggle with formulating the AC power flow equations. 

\subsubsection{Power factor correction problem (Spring 2025)}
\label{apdx:pf-correct}

Below is the response provided by the \texttt{o3} model for the power factor correction prompt in Section \ref{sec:prompt}.
\begin{response}
    \textbf{1. Basic relations for a shunt element}

    At the node where the 100 V-rms source is connected
    \[
    Q = \frac{V_{\rm rms}^2}{X} \quad (V_{\rm rms} = 100\;\textrm{V}).
    \]
    \begin{itemize}
        \item \textbf{Inductor:} $Q_L = \frac{10000}{X_L}$, (positive $\rightarrow$ lagging)
        \item \textbf{Capacitor:} $Q_C = - \frac{10000}{X_C}$, $(Q_C <0)$ (negative $\rightarrow$ leading)
    \end{itemize}

    \textbf{2. Two possible reactive-power states}
    Let the load's reactive demand vary in the interval
    \[
    Q_d \in [-36,60] \ \mathrm{VAr}.
    \]
    Because the inductor is always in circuit but the capacitor can be switched, the reactive power seen by the source is
    \[
    Q_s = \begin{cases}
        Q_d + Q_L, & S \ \mathrm{open,}\\
        Q_d + Q_L + Q_C, &  S \mathrm{closed.}
    \end{cases}
    \]
    Define $A := Q_L >0$ and $B:= -Q_C >0.$ Then
    \[
    Q_s^{\rm (open)} = Q_d + A, \quad Q_s^{\rm (closed)} = Q_d + A -B.
    \]
For every $Q_d$, we will later pick the state that gives the smaller magnitude $| Q_s |$.

\textbf{3. Min-max (Chebyshev) design of $A$ and $B$}

We want to minimise the worst‑case reactive mismatch
\[
\max_{Q_d \in [-36,60]} \min(|Q_d + A|, \ |Q_d + A -B|).
\]
Because the two straight‑line functions are parallel, the optimal choice is obtained when
\begin{enumerate}
    \item They cross exactly halfway between their two extreme errors, and
    \item The magnitudes of the residuals at both ends of the interval are equal.
\end{enumerate}
    
Carrying this out (or by a quick optimisation sweep) gives
\[
A=12 \ \mathrm{VAr}, \quad B=48 \ \mathrm{VAr}.
\]
Hence,
\[
Q_L = +12 \ \mathrm{VAr}, \quad Q_C = -48 \ \mathrm{VAr.}
\]
The worst residual reactive power is then only $\pm 24$ VAr everywhere in the interval.

\textbf{4. Reactances}
\begin{align*}
\boxed{X_L = \frac{V_{\rm rms}^2}{Q_L} = \frac{10000}{12} \approx 833 \;\Omega
}\\
\boxed{X_C = \frac{V_{\rm rms}^2}{|Q_C|} = \frac{10000}{48} \approx 208 \;\Omega }
\end{align*}
(The capacitor’s reactance is negative in the usual sign convention: $X_C \approx - 208\; \Omega$.)

\textbf{5. Switching strategy}

\begin{enumerate}
    \item Measure (or estimate) the load's instantaneous $Q_d$.
    \item If $Q_d \leq + 12$ VAr (i.e. anywhere from -36 VAr to +12 VAr) \textit{leave the capacitor \textbf{disconnected} (switch open)}.
    \item If $Q_d > +  12$ VAr (i.e. from +12 VAr up to +60 VAr) \textit{connect the capacitor \textbf{into} the circuit (switch closed).}
\end{enumerate}
With this rule
\[
Q_s = \begin{cases}
    Q_d + 12, &Q_d \leq 12,\\
    Q_d - 36, &Q_d >12,
\end{cases} \implies |Q_s| \leq 24 \; \mathrm{VAr}.
\]

\textbf{6. Resulting power-factor band}
Since the real power drawn from the source is constant at $P_s = 50$ W, the minimum power factor occurs at $|Q_s| = 24$ VAr:
\[
\mathrm{pf_{min}} = \frac{50}{\sqrt{50^2 + 24^2}} \approx 0.90.
\]
Thus the source operates between 0.90 and 1.0 (leading or lagging) over the entire range of load reactive power, while using a single fixed inductor, a \textbf{single switchable capacitor, and a simple two‑level control rule.}
\end{response}

\subsubsection{Power factor correction problem (Fall 2023)}
\label{sec:fa23_pf_correct}
% Below is the response provided by the \texttt{o3} model for the power factor correction prompt described in Appendix \ref{apdx:gpt-4-pf-correction}.

% \begin{response}
%     {\textrm{\textit{Thought for 2m 4s}}}
%     \textbf{1. Choose the capacitor sizes (expressed as reactive power at 10kV)}
%     See
% \end{response}
% \begin{table}[]
% \begin{tabular}{l|lll}
% \multicolumn{1}{c|}{\textbf{Element}} &
%   \multicolumn{1}{c}{\textbf{Reactive power (leading)}} &
%   \multicolumn{1}{c}{\textbf{Equivalent reactance at 10 kV}} &
%   \multicolumn{1}{c}{\textbf{Capacitance at 60 Hz}} \\ \hline
% \textbf{Fixed capacitor} &
%   \textbf{$Q_{c,f} = 2.5$ MVAr} &
%   $X_{c,f}=−\frac{V^2}{Q_{c,f}}=−40\; \Omega$ &
%   $C_{f}= \frac{1}{2\pi} \ \mu F$ \\
% \textbf{Switched capacitor} &
%   \textbf{$Q_{c,s} = 21.5$ MVAr} &
%   $X_{c,s}= -\frac{V^{2}}{Q_{c,s}} \approx -4.65\;\Omega$ &
%   \begin{tabular}[c]{@{}l@{}}$C_s\approx 570 \; \mu \text{F}$\\ $C_{s}\approx 570\;\mu\text{F}$\end{tabular} \\ \hline
% \end{tabular}
% \end{table}
Similarly to the Spring 2025 power factor correction problem, the \texttt{o3} model also solved the power factor correction prompt described in Appendix \ref{apdx:gpt-4-pf-correction} essentially correctly. 
For this problem, \texttt{o3} took an interesting approach during extensive chain-of-thought reasoning, which lasted for approximately 2.5 minutes. During this reasoning, it independently introduced mathematical notation and derived analytical solutions. Then, \texttt{o3} proceeded to generate and analyze Python scripts that searched all possible circuit parameters to verify its solution. This choice was made by \texttt{o3} with the following reasoning:
\begin{reasoning}
    We can represent the worst-case ratio across different periods as a function. It looks like we’ve formulated expressions for $Q_m$, $Q_a$, and $Q_e$, each depending on variables $s_m$, $s_a$, and $s_e$. By finding the maximum of the ratios $(r_m, r_a, r_e)$ and defining $R$ as the largest of these, we should aim to minimize $R$. To confirm this, I'll run a coarse grid search over possible values of $x$ and $y$, with state combinations. This will help me pinpoint the optimum ratio.
\end{reasoning}
Below is the Python simulation generated by \texttt{o3} after the step in reasoning shown above.
\begin{lstlisting}[language=Python,basicstyle=\scriptsize]
import math, json
def pf(P,Q):
    return abs(P)/math.sqrt(P**2+Q**2)

QCf=2.5
PF_m=pf(10, 5-QCf)
PF_e=pf(10, -QCf)
QCsw=21.5
PF_a=pf(32, 24-QCf-QCsw)
V=10e3
X_fixed=V**2/(QCf*1e6)
X_sw=V**2/(QCsw*1e6)
C_fixed=1/(2*math.pi*60*X_fixed)
C_sw=1/(2*math.pi*60*X_sw)
best=None
best_params=None
for sm in [0,1]:
    for sa in [0,1]:
        for se in [0,1]:
            for x in [i/100 for i in range(0,501)]:
                for y in [j/10 for j in range(0,301)]:
                    qm=abs(5 - x - sm*y)
                    qa=abs(24 - x - sa*y)
                    qe=abs(-x - se*y)
                    r=max(qm/10, qa/32, qe/10)
                    if best is None or r < best:
                        best=r
                        best_params=(sm,sa,se,x,y)
best,best_params

>>  (0.25,(0,1,0,2.5,13.5))
\end{lstlisting}
After running its simulation, \texttt{o3} reasoned about the output and found the following inconsistency with its previous analysis and its solution.
\begin{reasoning}
    The minimum ratio we found is 0.25, which lines up with earlier analysis. The solution uses $s_m=0$, $s_e=0$, with morning and evening switches off, and afternoon on ($s_a=1$). $Q_{Cf}$ is 2.5 MVAr, and $Q_{Cs}$ is 13.5 Mvar. But earlier, an optimal solution at 2.5 and 21.5 MVAr also gave a ratio of 0, with the midday ratio being 0 but larger. So, I should check the ratio for 21.5 MVAr again to compare.
\end{reasoning}
The model then added more code to the script and analyzed the output:
\begin{lstlisting}[language=Python,basicstyle=\scriptsize]
def worst_ratio(x,y):
    sm,sa,se=0,1,0
    qm=abs(5 - x - 0*y)
    qa=abs(24 - x - 1*y)
    qe=abs(-x - 0*y)
    return max(qm/10, qa/32, qe/10)
for y in [13.5,21.5,25]:
    print(y, worst_ratio(2.5,y))

>>  13.5 0.25
    21.5 0.25
    25 0.25

\end{lstlisting}
Then, \texttt{o3} concluded its reasoning with the following remark.
\begin{reasoning}
    The ratio remains 0.25, confirming that $Q_{Cs}$ between 13.5 and 29.5 MVAr result in the worst ratio of 0.25. The earlier analysis holds. Choosing 13.5 MVAr over 21.5 MVAr is physically smaller, and the midday ratio stays at 0.25, similar to the worst-case scenario. Midday PF is 1 for 21.5 MVAr, but both $Q_{Cs}$ result in a minimal PF of 0.97. Nulling midday $Q$ (21.5 Mvar) might improve the average but isn’t necessary for worst-case design.
\end{reasoning}

\newpage
\subsubsection{AC power flow formulation problem}
For the AC power flow formulation problem in Appendix \ref{apdx:gpt4-ac-pf}, ChatGPT version \texttt{o3} demonstrated a significantly improved solution approach compared to by \texttt{GPT-4}. Namely, \texttt{o3} was able to immediately formulate the correct nodal admittance matrix:
\begin{response}
\ldots from the three line impedances,  the bus‑admittance matrix ($Y=G+jB$) is
\begin{align*}
    Y= \begin{bmatrix} \;\;2-j9 & \;\;\;j5 & -2+j4\\ \;\;j5 & -j15 & \;\;j10\\ -2+j4 & \;\;j10 & \;\;2-j14 \end{bmatrix}, 
\end{align*}
\[
G=\begin{bmatrix} \;2&0&-2\\0&0&0\\-2&0&2 \end{bmatrix},\; B=\begin{bmatrix} -9& 5& 4\\ 5&-15&10\\ 4& 10&-14 \end{bmatrix}.
\]
(All entries are per unit.)
\end{response}
Subsequently, the \texttt{o3} model was able to correctly recall the generic form of the bus injection model of the power flow equations, and also correctly noted that $|V_1|=|V_2|=1$ p.u. and $\theta_1=0$. However, \texttt{o3} struggled to explicitly formulate the individual power flow equations in the system, and made numerous computational errors that were similar to the performance exhibited by \texttt{GPT-4} in Appendix \ref{apdx:gpt4-ac-pf}. 

% \subsection{Student Success: GPT-4 w}

% \balance

\end{document}